\documentstyle[12pt]{article}
\input epsf

\def\LG{Lan\-dau-Ginz\-burg\ }
\def\BW{Boltz\-mann weights\ }
\def\smvert{\vrule height7.5pt depth-.5pt width.2pt}
\def\IC{\relax\,\hbox{$\smvert \kern-.23em{\rm C}$}}
\def\IR{\relax{\rm I\kern-.18em R}}
\font\sanse=cmss12
\def\half{{1 \over 2}}
\def\ZZ{\relax{\hbox{\sanse Z\kern-.42em Z}}}
\def\scf{{\cal F}}
\def\Gminus{G_{-{1 \over 2}}^-}
\def\Gplus{G_{-{1 \over 2}}^+}

\def\annp#1{{\sl Ann. \ Phys.}\ {\bf #1\/}}
\def\cmp#1{{\sl Commun.\ Math. \ Phys.} \ {\bf #1\/}}
\def\nup#1{{\sl Nucl.\ Phys.} \ {\bf B#1\/}}
\def\plt#1{{\sl Phys.\ Lett.}\ {\bf #1\/}}
\def\ijmp#1{{\sl Int. \ J.\ Mod. \ Phys.}\ {\bf A#1\/}}
\def\ijmpb#1{{\sl Int. \ J.\ Mod. \ Phys.}\ {\bf B#1\/}}
\def\jsp#1{{\sl J. \ Stat. \ Phys.}\ {\bf #1\/}}

\def\mpl#1{{\sl Mod.\ Phys.\ Lett.} \   {\bf A#1}\ }
\def\jphys#1{{\sl J. \ Phys.}\ {\bf A#1\/}}

\begin{document}

\hsize37truepc\vsize61truepc
\hoffset=-.5truein\voffset=-0.8truein
\setlength{\baselineskip}{17pt plus 1pt minus 1pt}
\setlength{\textheight}{22.5cm}


\title{Lattice models and $N=2$ supersymmetry}

\author{H.~Saleur and N.P.~Warner  \vspace{0.5cm}
\\
Physics Department \\
University of Southern California \\
University Park \\
Los Angeles, CA 90089-0484 \\
U.S.A. }
\date{ }
\maketitle

\vspace{1cm}
\begin{abstract}
We review the construction of exactly solvable lattice models whose
continuum limits are $N=2$ supersymmetric models.  Both critical
and off-critical models are discussed.  The approach we take is
to first find lattice models with natural topological sectors, and
then identify the continuum limits of these sectors with
topologically twisted $N=2$ supersymmetric field theories. From this,
we then describe how to recover the complete lattice versions of the
$N=2$ supersymmetric field theories.
We discuss a number of simple physical
examples and we describe how to construct a broad class of models.
We also give a brief review of the scattering matrices for the
excitations of these models.
\end{abstract}
\vspace{1cm}
\hspace{11mm}USC-93/026 \hfill \\
\vspace{1mm}
\hspace{11mm}hep-th/9311138  \hfill \\
\vspace{1mm}
\hspace{10mm}November, 1993 \hfill \\
\thispagestyle{empty}

\newpage
\setcounter{page}{1}

\section{Introduction}

\index{Conformal field theory!perturbed}
\index{Conformal field theory!integrable models}
\index{Supersymmetry!$N=2$}

$N=2$ supersymmetric theories in two dimensions are in some ways much
simpler
than their non-supersymmetric kin.  This is essentially because the
$N=2$
supersymmetry imples the presence of a topological sector for which
semi-classical
analysis yields exact quantum results.   (In the language of
supersymmetry,
the ``$F$-terms'' are not renormalized.)  This does not mean that the
model is
semi-classically rigid, but only that a key (topological)
subsector of the theory is
semi-classically determined: the complete theory is very rich and has
all of
the complexity of a non-supersymmetric theory.  The topological
subsector has
thus provided a ``bridgehead'' from which many of the
non-trivial quantum aspects of the model can be explored, and usually
with
greater facility, and often in more detail than is possible for
non-supersymmetric theories. (For reviews, see
\cite{Martrev,Wrevs,Vafrev}.)
For example, $N=2$ \LG theories have a
superpotential that is exact.  For $N=2$ superconformal models this
means
that the operator algebra and the anomalous dimensions (conformal
weights)
of the \LG fields are trivially computable \cite{VWMLG,LVW}.  On the
other
hand,
it is still somewhat unclear as to how much of the rest of the
theory, and in
particular, how much of the complete operator content
is determined by the \LG potential.  \index{Landau-Ginzburg}
Indicative results are known:
For example, the ADE classification of modular invariants of $N=2$
minimal
models collapses to the ADE classification of modality zero
singularities
\cite{CIZ,VWMLG}.  Ramond sector characters can be obtained from the
\LG
potential by computing the elliptic genus
\cite{EWell,PDOASY,TKYYSKY,MHenn}.

For $N=2$ supersymmetric quantum
integrable field theories (QIFT's \footnote{In this review, conformal
field theory will be abbreviated as CFT, quantum field theory as QFT,
and
quantum integrable field theory as QIFT.}) the \LG description
leads to transparent analysis of the soliton structure
\cite{FMVW,PFKI,ALDNNW}.
In $N=2$ QIFT's, and even non-integrable $N=2$ QFT's,
differential equations can also be obtained
for some of the scaling functions \cite{CecVaf}.  For the $N=2$
QIFT's these
scaling functions can also be obtained from the thermodynamic Bethe
ansatz,
but instead of differential equations, one obtains
complicated integral equations \cite{CFIV}.

Over the last decade it has also become evident that many of the
structures of
quantum integrable field theories have analogues in exactly solvable
lattice
models
(see, for example  \cite{VPasq,VPHS,BNI,DFSZI,CoulombG}).   Using
this
technology,
lattice models have been constructed in which the continuum limits
give rise to
many
of the non-supersymmetric conformal field theories.
It is therefore natural to expect that there
should be exactly solvable lattice models whose continuum limits are
$N=2$ supersymmetric QIFT's.  We shall henceforth refer to such
lattice
models as $N=2$ lattice models\footnote{This nomenclature is somewhat
misleading
in that it suggests that the supersymmetry is realized on the
lattice,
whereas it still remains unclear whether this can be done in most of
the
$N=2$ lattice models thus far constructed.}.  It is also to be hoped
that
such lattice regularizations may have especially simple features.
If so, they may in turn help us to understand some mysterious issues
like the appearance of branching functions in local height
probabilities
\cite{Kyoto}.  There are also practical reasons for constructing
$N=2$ lattice
models.  The simplicity of $N=2$ QFT's has led to more progress than
in any
other kind of field theory.  It is especially attractive to try to
compare some of the advanced results with real or computer
experiments. To do
so, the construction of a ``physically reasonable'' lattice model
whose
continuum
limit is the $N=2$ QFT of interest is a natural way to proceed.  We
also hope
that the $N=2$  lattice models will be ``more universal'' than their
non-supersymmetric counterparts.  Indeed, consider a general
non-supersymmetric
model, and suppose that we do not impose the requirement of exact
solvability.
Since there are generically very many relevant operators in such a
model,
the corresponding coupling constants would need to be very finely
tuned
in order to
find a particular second order phase transition and hence a
particular
conformal field theory. Thus one of the side-effects of exact
solvability
is to automatically make such a fine-tuning of the couplings.
The $N=2$ lattice models have a further special, and robust,
identifying feature: the presence of a topological sector.  Thus, not
only
can one construct them by imposing exact solvability, but one can
also identify them by virtue of their topological sector.
In fact, we will argue in this review that any ``trivial''
statistical
mechanics model will probably give rise to a non-trivial $N=2$
lattice model,
and thus
the $N=2$ models will be of relevance in a large variety of
situations.
For instance the polymer and  percolation problems, the statistics of
Bloch
walls in
high temperature Ising model, Brownian motion, are all described by
$N=2$
supersymmetric theories.

The investigation of lattice models associated with supersymmetric
QFT's has a
rather long history. To our knowledge\footnote{We apologize for any
reference
missing in
this short history.} it started with the study of $N=1$ supersymmetry
in the tricritical Ising model \cite{FQS,KMS,ZQiu} and the fully
frustrated
XY-model
\cite{OFoda} (the latter related to Josephson junctions arrays).

Then $N=1$ supersymmetry
together with $N=2$ supersymmetry were observed at some special
points
of the Ashkin-Teller model and the six-vertex (or XXZ) model
\cite{SKY,SKYHBZ,Bonn}. By analysis of torus partition functions
it was  easy to identify more generally special points of
the higher spin  $U_q(SU(2))$ vertex models
\cite{DFSZ,SKYI} that were $N=2$ supersymmetric. A physical
explanation of this
came later \cite{HSAA}. In the simplest case of the six-vertex model
it leads
to
$N=2$ supersymmetry in the percolation problem. Similarly, by
analysis of the
related Izergin-Korepin ($O(n)$) model \cite{AGIVEK,WBNN}, another
family of lattice models with $N=2$ supersymmetry was found (or
conjectured).
The simplest of this family leads to $N=2$ supersymmetry
in the self-avoiding walk (polymer) problem. This turns out to be a
very
favorable case for comparison with real and numerical experiments
\cite{PFHS}.
More complete study of the $SU(2)$
based models was carried out in \cite{CGGS}. Recently, it has been
shown how
one can
go beyond models based upon $SU(2)$.  That is, $N=2$ lattice models
based on
any simply-laced Lie algebra have been constructed by using ``partial
restriction'' of
modified solid-on-solid (SOS) models, or equivalently by twisting and
performing
{\it partial} quantum group truncation of the corresponding vertex
models. The
continuum limits of these lattice models are the $N=2$ superconformal
coset
models
based on hermitian symmetric space \cite{MNW,ZMa,DNNW}.

At the present time, there is a respectable number of known $N=2$
lattice
models.
They resemble in some respects the corresponding QFT, although the
structures
are
not yet as closely linked  as one would wish. They have already led
to
comparison
with real and numerical experiments, and very good agreement has been
found.
The
purpose of this review is to describe in some detail
the current state of knowledge and to point to new directions of
research.

The first section is rather qualitative and collects
observations  about the expected structure of $N=2$ lattice models. A
variety
of simple examples is worked out, starting from simple geometrical
ideas and
introducing the fundamental tools that will
be used extensively later.  One of the basic themes underlying the
$N=2$
lattice constructions is the role of the ``trivial'' topological
sector.  This
will
be used directly in later sections to construct and analyse the $N=2$
lattice
models
based on general Lie algebras.  In the fourth section we will briefly
discuss the scattering matrices of excitations in the $N=2$ lattice
models and
their
continuum limits.   In section five we discuss further physically
interesting
aspects
of $N=2$ lattice models, and in the last section we conclude by
summarizing
what
we believe are the important open problems in the subject.

\section{Features of  $N=2$ lattice models.}

Our purpose in this section is to study a number of closely related
and simple
models that exhibit the basic features of the $N=2$ lattice models.
Our aim is
to try
to give some intuitive understanding of $N=2$ lattice models, to show
how one
can analyse
their various features, and ultimately to evolve a strategy for
finding such
models.

\subsection{The clues from the topological sector}

\index{Topological theories}
If the theory has $N=2$ supersymmetry then, as mentionned earlier,
there is
a topological sector, with  the following characteristics.  The
topological
sector
of the $N=2$ QFT consists
of operators that are annihilated by two of the four supercharges.
The
topological
physical\footnote{The definition of  ``physical'' for the lattice
model can be
quite
different that of the field theory.}
states  consist of only the ground states of the non-topological
$N=2$
lattice model.  The operators of the topological model are order
parameters of
the non-topological sector, and the correlation functions of these
operators
are
constant in the topological sector.

The presence of some sort of  ``topological'' sector is one of the
main
attributes of
an $N=2$ lattice model, and because of the links between the
topological and
non-topological sectors in the continuum,
the identification of the lattice topological sector proves an
important tool
in the
construction of the complete $N=2$ lattice model.  More precisely,
the idea is
to study
representations of the lattice algebras and lattice symmetries (such
as
Temperly-Lieb
and quantum group representaions).  For certain choices of parameters
some of
these
will be very simple or trivial, and the corresponding lattice models
will
be ``topological.''  The idea is to define the topological model
precisely, fix
the
parameters to their topological values, but then modify the choice of
representations or
modify the choice of lattice variables in such a manner that the
$N=2$ lattice
model emerges
from its topological limit.  We will describe two ``dual'' methods
for
accomplishing this.
The first relies on the fact that a lattice model can
have very different properties when it is described by different sets
of observables that are not mutually local.  One thus obtains some of
the $N=2$
lattice models by  considering a ``trivial'' model where the given
lattice
variables
do not interact, and then one makes a ``change of variables'' to new
set of
geometrical
observables that
have highly non-trivial properties.  The non-interacting observables
describe
the
topological model, while the new observables describe the non-trivial
$N=2$
lattice
model.  The second method is to consider the restriction process in
solid-on-solid
(SOS) models (or quantum group truncation in the equivalent vertex
formulation), and
start with a restriction that results in a ``frozen'' or completely
rigid
topological
model.  One then ``gets something from nothing'' by relaxing the
restrictions
imposed on
the SOS model, while at the same time one ``topologically untwists''
the \BW to
avoid
singularities in the transfer matrix.
It turns out that the first method is {\it a priori} a little more
intuitive,
while the
second method is much easier to generalize.  We therefore start by
describing a
simple
example of the first procedure and then trace a connection to the
second
method.

\subsection{Examples of lattice topological sectors}

\subsubsection{Percolation: an example of a geometrical model}

\index{Topological lattice models}
\index{Percolation}
Consider a square lattice and put on every edge a variable $\sigma$
that can
take two possible values, $0$ or $1$.  Suppose that the edge
variables do not
interact with each other, but put the whole system in a magnetic
field.
Since the spins do not interact,  all correlation functions are
trivially constant (or, at least are delta functions).
This appears to be a completely trivial model and the partition
function is
${\cal Z}=\left(1+e^{-H}\right)^{N_E}$, where ${N_E}$ is the
number of edges.    We now pass to a new set
of observables (called geometrical observables in what follows):
Call  edges with  variable $\sigma=0$ ``empty'' and
edges with variable $\sigma=1$ ``occupied'' (we represent occupied
edges with
heavy
lines in figure 1) and consider the geometrical properties of
clusters, that
is, of
the connected sets of occupied edges.  These cluster observables
are non-local with respect to the original $\sigma$ variables, and
the
``trivial''
model now describes the bond percolation problem.
Since the sum over variables $\sigma=0,1$ is unconstrained,
there is a probability of $p={e^{-H} \over 1+e^{-H}}$ of having an
occupied
edge,
and a probability of $p={1\over 1+e^{-H}}$ of having an empty edge.
The correlation functions of new variables, such as the probability
that two
vertices
belong to the same cluster, are non-trivial. The original trivial
model can,
of course,
be recovered by deciding to study only  local questions like ``is
this edge
occupied''
and forgetting about non local questions like ``are these two edges
part of the
same
cluster''.  As we will see, the distinction between these two
questions is
clearer
from a representation theory point of view, and easy to implement,
for
instance, in
a transfer matrix  formalism.

\begin{figure}[htb]
\epsfxsize = 6cm
\vbox{\vskip .8cm\hbox{\centerline{\epsffile{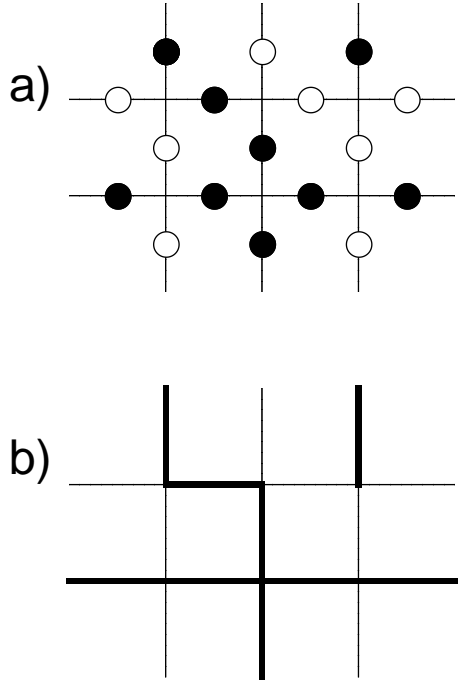}}}
\vskip .5cm \smallskip}
\caption{a) A configuration of ``up'' (black dot) or
``down'' (white dot) spins on the edges of the square lattice.
b) The corresponding configuration of
``occupied'' or ``empty'' edges.  }
\end{figure}

To determine the supersymmetry point of this model we consider it
upon a torus.  The new observables can be given various boundary
conditions: for instance, one can sum over non-contractible clusters
giving each some weight, or fugacity.  The presence of unbroken
supersymmetry
means that, in the thermodynamic limit,
the free energy per edge cannot depend upon the boundary
conditions on the torus.  Intuitively, if $p>1/2$  or $p<1/2$, one
will
find large clusters of occupied or unoccupied edges.  Thus
non-contractible
clusters will make significant contributions to the free energy, and
thus
the latter will depend upon the boundary conditions.  However, when
one has
$p=1/2$, or $H=0$, one is at the percolation point,  a critical
point where the geometrical correlation functions decay algebraically
and there
is a vanishing fraction of big clusters of occupied or unoccupied
edges.
The partition function is ${\cal Z}=2^{N_E}$.
This lattice partition function is, as usual, defined up to
a non-universal term of the form $e^{fN_E}$.  In the continuum, the
partition
function simply counts the number of ``different observables'' of the
model. It
is rather reasonable that there is only one such observable  (beside
the identity): the variable $\sigma$. One thus expects $Z=2$.  This
is
also consistent with the identification of this percolation problem
as
an $N=2$ lattice model.  As we will see, the foregoing lattice
partition
function can be interpreted as the Witten index of the $N=2$ lattice
model.
We will soon present rather more direct evidence that for $p= 1/2$
this is
indeed
an $N=2$ lattice model.

\subsubsection{The dual point of view: the ``frozen'' Potts model}

\index{Potts model}
The foregoing non-trivial bond percolation problem can also be
described using
the
$Q$-state Potts model in the limit $Q \to 1$.  The general $Q$-state
Potts
model
is defined by putting a spin variable, $\sigma$, on each {\it vertex}
of the
square lattice.  These spin variables take values in $1,\ldots,Q$,
and the
lattice
partition function is
\begin{equation}
{\cal Z}_{\rm Potts} ~=~ \sum_{\{\sigma\}}\prod_{<i;j>}
{}~e^{K ~\delta(\sigma_i, \sigma_j)} \ ,
\label{pottspart}
\end{equation}
where the sum is over all spins and the product is over is over
nearest
neighbours.
The connection with the percolation problem
becomes evident upon making a high temperature expansion of the Potts
model:
The partition function can be written
\begin{equation}
Z_{\rm Potts} ~=~ {\rm tr}~\left( \prod_{<i;j>} \left\{ 1 ~+~ (e^K -
1)
\delta (\sigma_i, \sigma_j) \right\} \right) \ ,
\end{equation}
which can then be expanded graphically by connecting all vertices
with the same
spin
$\sigma$.  One then finds that
\begin{equation}
Z~=~ \sum_{\rm graphs} (e^K - 1)^{N_B} ~ Q^{N_C} \ ,
\end{equation}
where $N_B$ is the number of bonds in the graph, and $N_C$ is the
number of
clusters.    This high temperature expansion coincides with the bond
percolation problem only if the $Q^{N_C}$ term is equal to one, {\it
i.e.}
if and only if $Q=1$.  One must also make the following
identification:
\begin{equation}
e^K - 1 ~=~ e^{-H} \ .
\label{KHreln}
\end{equation}
However, if one considers the Potts model with $Q=1$, one finds that
it is a
trivial, completely frozen, ``topological'' model.  The partition
function
(\ref{pottspart}) collapses to ${\cal Z}_{\rm Potts} = e^{K N_E}$.
The bond
percolation
problem should really be viewed as the $Q \to 1$ limit of the Potts
model.  For
example, the derivatives of ${\cal Z}_{\rm Potts}$ at $Q=1$ describe
generating
functions
for the percolation problem \cite{KastFort}.  Similarly, derivatives
of
the Green functions correspond to geometrical correlation functions;
for
example the
spin-spin two point function of the Potts model yields the
probability that two
vertices
belong to the same cluster.

We therefore see that the percolation problem is closely associated
with a
frozen model.
To obtain the non-trivial model from the frozen model we need to find
models
that are ``close to the frozen model.''  The natural way of doing
this is to
use
an algebraic approach.

\subsubsection{Representations of the Temperley-Lieb algebra}

\index{Temperley-Lieb algebra}
To proceed further it is  convenient to think in algebraic terms and
turn to a transfer matrix formalism \cite{BlotNight}. Consider thus a
rectangle
of size $L\times T$ and propagation in the $\hat{y}$ direction.  We
define more precisely the geometrical observables in terms of
connectivities
as follows.  Number the vertices on a given time slice  by
$1,2,\ldots,L$,
running
from left to right. The space  of states ${\cal H}_L^{\rm geom}$ is
the set of all possible partitions of  $1,2,\ldots,L$.  In a
partition of $L$
at
time $t$, two vertices are grouped together if at time $t$ they are
connected
through a cluster that extends into their present and past.   For
instance,
for $L=4$ the partition $(13)(2)(4)$ correspond to the fact that
there is
connectivity between $1$ and $3$, while $2$ and $4$ are not currently
connected to any other vertex by connections through clusters in
their past
(see figure 2).  Introduce operators, $e_k$, that act on ${\cal
H}_L^{\rm
geom}$.
For odd subscripts, $e_{2j-1}$ transforms connectivities at time $t$
to
connectivities at time $t+1$ by inserting vertical ``occupied'' bonds
at all
but the $j^{\rm th}$ vertex.  Thus one has
$e_3|(13)(2)(4)>=|(13)(2)(4)>$, and
$e_1|(13)(2)(4)>=|(1)(2)(3)(4)>$.  For even subscripts, $e_{2j}$,
modifies
connectivities at the same time, and simply inserts a
horizontal edge between the $j^{\rm th}$ and $(j+1)^{\rm th}$
vertices.
For example, one has $e_2|(13)(2)(4)>=|(123)(4)>$.

\smallskip
\begin{figure}[htb]
\epsfxsize = 6cm
\vbox{\vskip .8cm\hbox{\centerline{\epsffile{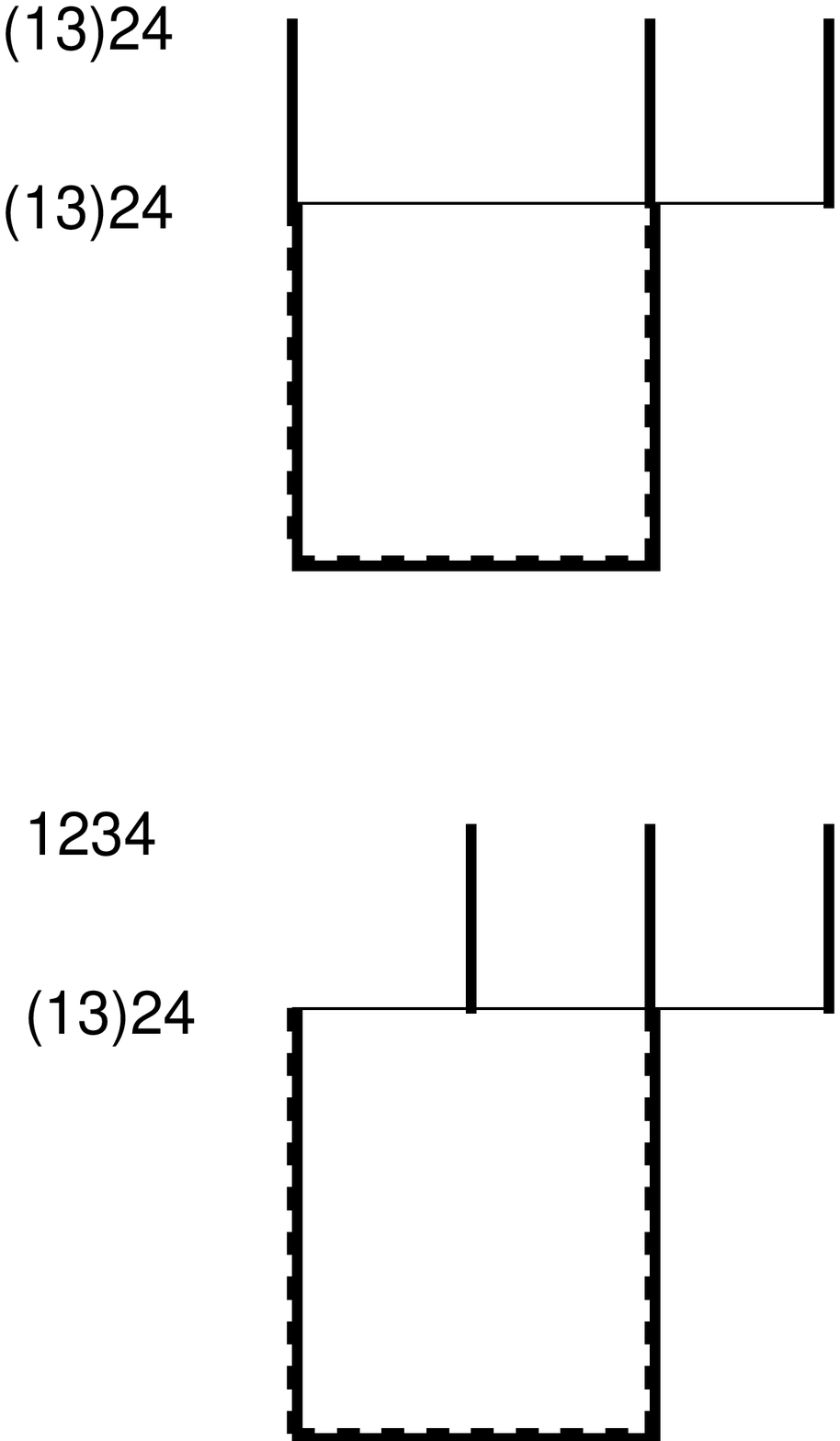}}}
\vskip .5cm \smallskip}
\caption{The partition $(13)24$ means that vertices 1 and
3 are connected through a path of occupied edges at previous times
(by convention past means
edges below $y=t$). Acting with $e_3$ corresponds to adding an
occupied
vertical
edge in all but the second column between $y=t$ and $y=t+1$ resulting
again in $(13)24$.  Acting with $e_1$ corresponds to adding an
occupied
vertical
edge in all but the first column between $y=t$ and $y=t+1$  and this
destroys the connection between vertices 1 and 3, resulting in the
partition
$1234$.}
\end{figure}

These operators satisfy the relations
\begin{equation}
e_j^2=\sqrt{Q}e_j,\qquad e_j e_{j\pm 1}e_j=e_j,\qquad [e_j,e_k]=0,\
|j-k|\geq 2
\end{equation}
with $\sqrt{Q}=1$, and hence furnish a particular representation of
the
Temperley-Lieb algebra \cite{Baxter,PMar,WBNN}.
\index{Temperley-Lieb!representations}
The general form of this algebra can be realized in a number of
simple models, and, in particular,in the $Q$-state Potts model.

Define  ${\cal H}_{L}^{\rm Potts}$ to be the space of all possible
(Potts model) spin states on the $L$ vertices across a time slice of
the
lattice.
Introduce the following operators on  ${\cal H}_{L}^{\rm Potts}$:
\begin{equation}
\left(e_{2j}\right)_{\sigma,\sigma'}=\sqrt{Q}~ \prod_k ~
\delta(\sigma_k,\sigma_k')
{}~\delta(\sigma_j,\sigma_{j+1})\label{spinrep}
\end{equation}
and
\begin{equation}
\left(e_{2j-1}\right)_{\sigma,\sigma'}={1\over\sqrt{Q}}~\prod_{k\neq
j}~\delta(
\sigma_k,\sigma_k')\ . \label{spinrepI}
\end{equation}
The operators $e_j$ satisfy the Temperly-Lieb algebra for general
$Q$.
The transfer matrix of the Potts model may be written:
\begin{equation}
\tau=Q^{L/2}\prod_{j=1}^L X_{2j}\prod_{j=1}^L X_{2j-1}
\label{transfmata}
\end{equation}
where
\begin{equation}
X_{2j-1}={e^K-1\over\sqrt{Q}}+e_{2j-1},\
X_{2j}=1+{e^K-1\over\sqrt{Q}}e_{2j} \
{}.
\label{transfmatb}
\end{equation}
The decoupled spin model in the magnetic field is related to this by
setting
$Q=1$ and
$e^K = 1 + e^{-H}$ as in (\ref{KHreln}).  The transfer matrix of the
bond
percolation
problem is also given by (\ref{transfmata}), (\ref{transfmatb}), but
with
the Temperley-Lieb representation given at the beginning of this
subsection.
More generally
the transfer matrices of the vertex and RSOS models that will soon be
described, also read as
in (\ref{transfmata}) and (\ref{transfmatb}) but with their own
Temperley-Lieb
representation.

On a torus, the algebraic structure  is slightly more complicated due
to the
additional edge between the first and $L^{\rm th}$ vertex. The
correct algebra
is now the periodic Temperley-Lieb algebra. The various possible
weights
associated
to non-contractible clusters and the various natural  geometrical
sectors
have precise meaning in terms of traces and  representation theory
\cite{MartSal}.

\subsection{Unfreezing frozen models}

We believe that the foregoing features are generic for an $N=2$
lattice model.
Such a model
should be closely related to two trivial models: one rigid or frozen,
and the
other some kind of high temperature dual in which the lattice
variables are
decoupled
from one another.  In between these extremes is the Hilbert space of
the $N=2$
lattice model.
Common to all of these models is the representation theory of some
underlying
lattice algebra. The problem of making the $N=2$ lattice model is to
find the
proper
observables in the decoupled model, or to find the proper
``unfreezing'' of the
frozen model.
As we will discuss, this may all be thought  of in terms of choices
of the
representations of the lattice algebra.  We will also find that while
the
``decoupled''
description of the model is more intuitive, it is the ``unfreezing''
process
that is
easiest to implement more generally.   To define more precisely the
unfreezing
procedure, we first want to describe two other representations of the
Temperley-Lieb
algebra, and their frozen limits.

\subsubsection{Yet more frozen models}

\index{RSOS models}
Consider the representation of the
Temperley-Lieb algebra provided by restricted solid on solid (RSOS)
models.  (We refer to more general extensions of such models as
interaction-round-a-face
(IRF) models.)  These representations have $\sqrt{Q}=2\cos\pi/(m+1)$,
where $m$
is an integer.
The model is defined by introducing heights, $\ell =1,\ldots,m$, on
the
vertices and faces
of the original square lattice (figure 3).
Represent these heights as
the nodes of the Dynkin diagram of $A_m$. The transfer matrix now
acts on a
configuration space ${\cal H}_{2L}^{\rm RSOS}$ whose basis is given
by elements
$\{\ell_j, j =1,\dots,L\}$ with the constraint that neighbouring
vertices on
the
lattice carry heights that are neighbours on the $A_m$ diagram.  Let
$v_{\ell}$
be the components of the Perron-Frobenius eigenvector of the Cartan
matrix of
$A_m$
(or of the incidence matrix of the Dynkin diagram).  That is, $v_\ell
= sin(\pi
\ell
/(m+1))$.  The representation of the Temperly-Lieb algebra is then
given by:
\begin{equation}
\left(e_{j}\right)_{\ell\ell'} ~=~ \delta(\ell_j,\ell_{j+2})
{}~{\left(v_{\ell_{j+1}}
{}~v_{\ell'_{j+1}}\right)^{1/2}\over v_{\ell_j}}~\prod_{k\neq j+1}
\delta (\ell_j,\ell'_j) \ .
\end{equation}
The RSOS transfer matrix is then given by using this in
(\ref{transfmata}) and (\ref{transfmatb}).  To get the representation
with
$Q=1$, one
simply takes $m+1=3$.  The model is rigid, with heights alternating
between the
value 1 on one sublattice, and the value 2 on the other.  This model
thus has
two states,
depending upon which sublattice takes the value 1 and which takes the
value 2.

\begin{figure}[htb]
\epsfxsize = 6cm
\vbox{\vskip .8cm\hbox{\centerline{\epsffile{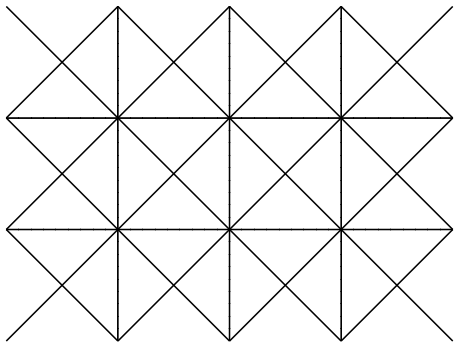}}}
\vskip .5cm \smallskip}
\caption{The RSOS variables are defined on the vertices of the
diagonal
lattice, that is the vertices and the faces of the original square
lattice.}
\end{figure}

This model has some form of duality with the original non-interaction
spin
model.
The non-interacting spin model can be considered as the
$T\rightarrow\infty$
limit of
an Ising model on the medial lattice. The dual \cite{Baxter} is an
Ising model
defined
on the faces and vertices of the original lattice and at $T=0$, which
is
completely ``frozen,'' and has exactly two states.

More importantly for our current purposes, this model has an
equivalent vertex
formulation whose continuum limit is a Gaussian model.  This
will finally yield more direct evidence of the supersymmetry.

\subsubsection{The vertex model and its continuum limits}

\index{Vertex models}
\index{Six-vertex model}
We now introduce our final representation
of the Temperley-Lieb algebra: the one  provided by the six-vertex
model.
Introduce spin variables $\pm$ (usually represented by arrows) on the
edges of
the
medial graph of the original square lattice (figure 4).   Think of
these spin
variables as being basis vectors in $\IC^2$.  The configuration space
of
the transfer matrix is now ${\cal H}_{2L}^{\rm vertex}=\left(
\IC^2\right)^{2L}$.  To define the operators $e_j$, introduce $4
\times 4$
matrices $E_{ij,kl}$, where $i,j,k,l$ are either $+$ or $-$, by
setting all the
entries to zero, except the $ij$,$kl$ entry which is set equal to
$1$.
Then $e_j$ is equal to the identity in all but the
$j^{\rm th}$ and $(j+1)^{\rm th}$ copies of $\IC^2$, where it is
given by
\begin{equation}
e_j=q^{-1}E_{+-,+-}+ qE_{-+,-+}-E_{+-,-+}-E_{-+,+-} \ .
\end{equation}
The parameter $Q$ of the Potts model is related to the parameter $q$
of the
six-vertex model by:
\begin{equation}
\sqrt{Q} ~=~ q+q^{-1} \ ,
\end{equation}
and so the supersymmetric model corresponds to setting
$q=\exp(i\pi/3)$.

\begin{figure}[htb]
\epsfxsize = 6cm
\vbox{\vskip .8cm\hbox{\centerline{\epsffile{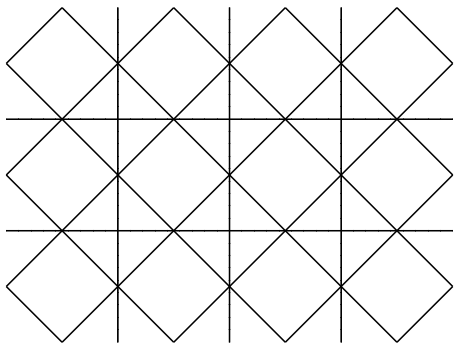}}}
\vskip .5cm \smallskip}
\caption{The six-vertex arrows are defined on the edges of the
diagonal
(medial) lattice. They
are also interpreted as domain walls for a SOS model whose heights
are
 defined on the vertices and faces of the square lattice (see figure
3).}
\end{figure}

The continuum limit of this model is a Gaussian model
\cite{PLKACB,BNI,DFSZI}.  \index{Gaussian model}
More precisely, introduce variables $\phi$ on the vertices
of the original lattice and in the middle of its faces, with values
defined
recursively by $\phi_u=\phi_d\pm 1$, that is,
the value of $\phi$ above one edge of the medial equal to the value
under it
plus the value of the spin carried by the edge.  The dynamics of
these
variables is described by a Gaussian action at large distance, and in
the
conventions where the topological defects are not renormalized, the
action is
\begin{equation}
A~=~{g\over 4\pi}~ \int ~ (\partial\phi)^2
\end{equation}
with $g=2/3$. As a result, the partition function on a torus with
doubly
periodic boundary conditions,
\begin{equation}
{\cal Z}~=~\hbox{tr} \left(~\tau^{\rm vertex}~\right)^T \ ,
\label{vertpart}
\end{equation}
is given, in the continuum, by
\begin{equation}
Z~=~Z_G(3/2) \ ,\label{neqltwopt}
\end{equation}
where $Z_G$ is a Gaussian partition function.  That is,
\begin{equation}
Z_G(g)~=~{1\over\eta\bar{\eta}}~\sum_{e,m\in Z}~
p^{h_{em}/4}\bar{p}^{\bar{h}_{em}/4} \ ,
\end{equation}
where
\begin{equation}
h_{em}~=~{1\over 4}\left({e\over\sqrt{g}}+m\sqrt{g}\right)^2,\qquad
\bar{h}_{em}~=~{1\over 4}\left({e\over\sqrt{g}}-m\sqrt{g}\right)^2
\end{equation}
and $\eta(p)=p^{1/24}\prod (1-p^n)$ is the Dedekind eta function,
and the elliptic nome is given by $p=\exp -2\pi T/L$. Equation
(\ref{neqltwopt})
coincides with the partition function
of the first minimal (central charge $c=1$) $N=2$ superconformal
model
with projection on odd fermion number.\index{Coulomb gas}  The
holomorphic
generators of the $N=2$
superconformal algebra can be written (after normalizing the boson
appropriately):
\begin{equation}
\begin{array}{lr}
G^+(z) ~=~ e^{i \sqrt{3} \phi(z)}\ ;&  G^-(z) ~=~ e^{- i \sqrt{3}
\phi(z)}\ ;
\\
J(z) ~=~ {i \over \sqrt{3}} \partial \phi(z) \ ;&  T(z) ~=~
- \half (\partial \phi(z))^2 \ .
\end{array}
\end{equation}

If one used the trivial representation of the Temperley-Lieb algebra
given by
equations (\ref{spinrep}) and (\ref{spinrepI}), instead of the
representation
given above, it would lead to
\begin{equation}
{\cal Z}~=~\hbox{tr}\left(\tau^{\rm Potts}\right)^T~=~ 2^{N_E} \ .
\end{equation}
In the continuum this partition function is, up to an overall
normalization,
the
partition function in the Ramond sector with an insertion of
$(-1)^F$, where
$F$ is the fermion number\footnote{In the purely bosonic formulation
the
fermion number
is defined by $F = e^{\pi i (J_0 + \tilde J_0)}$ , where $J$ and
$\tilde J$ are
the
holomorphic and anti-holomorphic $U(1)$ currents.}.  It is therefore
constant
and corresponds to the Witten index of the model.  It may also be
thought of as
the
partition function of the continuum topological matter model.

Beside partition functions, various physical observables in the
percolation
problem can be conveniently studied in the $N=2$ supersymmetry
formalism.
Usually the critical percolation problem is considered as a $c=0$
CFT,
{\it i.e.} a twisted $N=2$ superconformal theory.
The physical observables of statistical mechanics are then the ones
that
are usually discarded as unphysical from string theory point of
view. The probability that two edges belong to the same cluster
corresponds to an operator with half-integer labels in the Kac table
of the
$c=0$ CFT. Only by turning to an $N=2$ formalism can its correlation
functions
be studied. This operator belongs in the $N=2$ formalism to the
sector ``in
between'' Ramond and Neveu-Schwarz, with spatial boundary conditions
twisted by
 $(\sqrt{-1})^F$. Its dimension in the twisted theory is $h={5\over
96}$.

\subsubsection{A practical use of the topological sector}

\index{Free energy}
The identification of a topological sector is not only useful for
supersymmetry
purposes. If supersymmetry is unbroken, the free energy of a
non-trivial
lattice model, like the six-vertex model with $q=e^{i\pi/3}$, can
be readily computed by  turning
to the topological frozen model. At the special supersymmetric
point, it is not necessary to perform a Bethe ansatz computation to
determine
$f$.   This will be discussed further in section 3.

\subsubsection{Quantum group truncation, restriction and freezing}

\index{Quantum groups!representations}
\index{Qunatum group truncation}
Because the six-vertex model (with free boundary conditions) has a
quantum
group
symmetry, one can use this to
reduce the Hilbert space of the model to obtain a new truncated
model.  In the
language of height  models this correspond to making the RSOS
restriction.
It may also be viewed as the  lattice version of the BRST reduction
of the
Gaussian
model to the  Virasoro minimal models \cite{Felder}.  The first step
is to take
$q$
to be a root of unity.  Here we will take $q=e^{i \pi/m+1}$, for some
integer
$m$.
One then replaces the traces of operators by Markov
traces\index{Markov trace},
that is, the
trace of an operator  ${\cal O}$  is defined by:
\begin{equation}
tr_{M} ({\cal O}) ~=~\hbox{tr} \left(~{\cal O} ~ q^{2H} ~\right) \ ,
\label{Markov}
\end{equation}
where $H$ is the Cartan subalgebra generator of the $U_q(SU(2))$
symmetry of
the
model \cite{VPHS,MATHFRIENDS}  For example, the partition function of
the
truncated
vertex model is defined by ${\cal Z} = tr_M(\tau^T)$.

To see how this truncation works,
consider the Hilbert space of the vertex model, and imagine
decomposing it into representations of the $U_q(SU(2))$ symmetry.
The effect
of the factor of $q^{2H}$ in (\ref{Markov}) is to weight the
contribution of
each
representation by its $q$-dimension.  In particular, this means that
only the
type
${\cal I\!I}$ representations contribute to the trace\footnote{The
other
representations are called type ${\cal I}$ and are those
representations that
are reducible, but indecomposable.  Such representations also have
vanishing
$q$-dimension \cite{VPHS}.}.  This means that we can restrict the
trace in
(\ref{Markov}) to a trace over type ${\cal I\!I}$ representations.
The result
of this truncation in the continuum limit is that the Gaussian model
becomes
the
minimal model with central charge $c = 1 -{6 \over{m(m+1)}}$.

\index{Qunatum group truncation}
\index{Vertex/IRF correspondence}
To relate the vertex model to a height model one makes a change of
basis in the
vertex model Hilbert space \cite{VPasq,VPasqb,VPHS,HSJBZ}.  At the
vertices and
faces
of the original lattice   one introduces positive integer heights,
$\ell$,
in the following manner.  One views the arrows of the vertex model as
defining
the
basis elements of the spin-$\half$  representation of $U_q(SU(2))$.
One starts
at one
side of the lattice with a fixed element of some spin-$j_0$
representation
of  $U_q(SU(2))$;  if one tensors this with an element of the
spin-$\half$
representation
then one obtains a combination of vectors in the spin-$(j_0+\half)$
and spin-$(j_0-\half)$ representations.
Perform this tensoring successively with the spin-$\half$ states on
the edges
across the
lattice.   After each tensoring with a spin-$\half$ state, associate
the spin
of the  resulting $U_q(SU(2))$ representation to the next vertex.  At
the same
time,
keep track of the total Cartan
subalgebra eigenvalue of the state.  A sequence of such spins across
the
lattice, along with  Cartan subalgebra eigenvalue is a new basis for
the vertex
model
Hilbert space (figure 5).
The new basis is related to the old by a huge collection of
Clebsch-Gordon coefficients.  The quantum group symmetry means that
the
transfer
matrix is independent of the Cartan subalgebra eigenvalue, and so we
may
discard it.
The heights, $\ell$, are then related to the spins, $j$, by
$\ell=2j+1$.

\begin{figure}[htb]
\epsfxsize = 4cm
\vbox{\vskip .8cm\hbox{\centerline{\epsffile{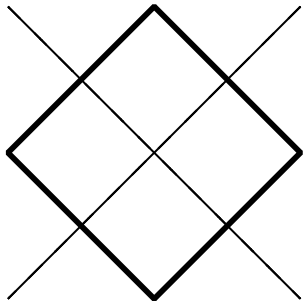}}}
\vskip .5cm \smallskip}
\caption{The geometry used in the vertex-RSOS transformation. Heights
sit on
the
vertices of the heavy square. The dotted edges carry the spins of the
six-vertex model. Heights
are viewed as $U_q(SU(2))$ highest weights and neighbours are
connected by
tensor
product with the fundamental representation.}
\end{figure}

\index{6j-symbols}
One can rewrite the transfer matrix, and indeed the individual
Boltzmann
weights,
in this new basis.
They are related to the vertex model \BW by quantum $6j$-symbols
\cite{VPasqb,HSJBZ}.
The effect of performing the quantum group truncation in the vertex
model is
equivalent to restricting the spins, $j$, of the representations on
each vertex
to
those of type ${\cal I\!I}$ representations. That is, for $q=e^{i
\pi/m+1}$,
one
restricts the heights to the range
$\ell = 1,2, \dots, m$.  In making the change of basis one finds that
because
the quantum $6j$-symbols vanish at certain strategic values of the
spins,
the the new \BW preserve
such a truncation.  One could naturally ask what would happen if one
made the
change
of basis but did not perform the truncation.  One would find that the
some
other
$6j$-symbols would vanish, and as a result some Boltzmann weights
would become
singular.
Thus, when $q$ is a
root of unity, this particular height formulation is necessarily the
restricted
(RSOS) model.

For $q=e^{i \pi/3}$, the only type ${\cal I\!I}$ representataions of
$U_q(SU(2)$ are
spin-$0$ and spin-$\half$.  Therefore, as has already been observed,
the
heights
in the RSOS model alternate between $1$ and $2$, and the model is
frozen.

{}From this perspective it is somewhat clearer what we must do in
order
to recover an $N=2$ model from a frozen model.  We must generalize a
process
that takes us back to the six-vertex model from the RSOS model.
Morally
speaking, this
should involve some form of releasing the height restriction, however
it is not
quite this simple since, as we mentionned above, doing this in the
RSOS model
will
result in both vanishing and singular Boltzmann weights.
The key point is to realize that the solution
to this problem of singular \BW is directly related to the process of
untwisting the energy-momentum tensor
of the continuum topological theory into the energy-momentum tensor
of the
$N=2$ superconformal theory.  In more concrete terms, the fact that
the
transfer
matrix in the six vertex model with free boundary conditions commutes
with the
quantum group means that the transfer matrix and the corresponding
spin-chain
hamiltonian necessarily contain special boundary terms \cite{VPHS}.
The
continuum limit of this hamiltonian is that of the topological
theory.
Thus to obtain the $N=2$ lattice model one has to twist the
transfer matrix of the vertex model with free boundary conditions so
as to
remove these
boundary terms \cite{MNW}.  Doing this also breaks the quantum group
symmetry
and thus
modifies the spectrum of the spin-chain hamiltonian \cite{ZMa}.  The
result is
the
pure Gaussian model described earlier.  An alternative way of
removing these
boundary terms  is simply to use periodic boundary conditions (this
was
implicitly used
earlier).  This works for the six vertex model, but not for its
generalizations,
in which we only want to break part of the quantum group symmetry.
We will discuss this more extensively
in the next section.

\subsection{Where is the supersymmetry on the lattice?}

As stressed above, one of the indications of supersymmetry in any
model is the existence of a trivial sector.  We have also seen that
the
$N=2$ superconformal generators can be explicitly constructed in the
continuum
limit.  One would, however, like to
see more direct evidence of the supersymmetry on the lattice.   For
example,
one would like to be able to exhibit lattice
quantities that reproduce the supersymmetry algebra. More generally,
one would
like to find some lattice fermion operators.

Unfortunately, although some progress has recently been made
\cite{WMKHS} in
the
identification of lattice quantities whose continuum limit is the
Virasoro
algebra,
the situation is still very unclear for supersymmetry generators. An
obvious
related difficulty  is that on the lattice we only see the symmetry
$U_q(SU(2))$
with $q=\exp(i\pi/3)$, while the continuum theory is characterized by
a pair
of quantum groups, the other one being $U_{q'}(SU(2))$,
$q'=\exp(i\pi/2)$, and
it is this
latter quantum group that has much to do with the supersymmetry
algebra. For
the
more general models of next section, we will again observe the
``wrong''
quantum group on the lattice.

\index{Polymers}
For the $c=1$ superconformal model there is, however, an $N=2$
lattice model
based
on polymers that exhibits a $U_{q'}(SU(2))$ symmetry. Although
no explicit realization of supersymmetry is known in this model, it
at least
allows a rather satisfying  identification of lattice  fermionic
degrees of
freedom.
Polymers, or self-avoiding walks, are better described  as the limit
of the $O(n)$ model \cite{DG} as $n\rightarrow 0$. The topological
model is
simply a model with no degrees of freedom at all, and with ${\cal
Z}=1$. The
non-local
geometrical observables are obtained by considering properties
of self-avoiding mutually avoiding walks on the lattice. Typical
correlation
functions at coupling $\beta$ have the form
\begin{equation}
\sum_N\beta^N\Omega_N \ ,
\label{cor}
\end{equation}
where the sum is taken over all self-avoiding walks connecting two
points (see figure 6), and $\Omega_N$, is the number of such walks of
length
$N$.
The critical point, where the correlators decay algebraically, occurs
when
$\beta=\beta_c$, where
$\beta_c^{-1}$ is the ``effective connectivity constant,''  which is
defined
for
large $N$ by  $\Omega_N\approx \beta_c^{-N},\ N>>1$.  The proper
algebraic setting for this model is $A_2^{(2)}$. It has however a
$U_q(SU(2))$ symmetry with $q=i$ \cite{FSZ}. A simple way of
identifying
fermionic degrees of freedom is to think of the zero weight given to
closed
loops as being the sum of statistical factors  $+1$, the
$-1$ corresponding to the loop carrying a bosonic or fermionic
variable. On a
torus,
the choice of antiperiodic boundary conditions for fermions gives a
weight
$1+1=2$
to some families of loops: this allows a simple recovery of the
various sectors
of the
$N=2$ theory from the lattice. This fermion interpretation allows
also
a transparent interpretation of the new index of \cite{CecVaf,CFIV}
in polymer
terms.

\begin{figure}[htb]
\epsfxsize = 6cm
\vbox{\vskip .8cm\hbox{\centerline{\epsffile{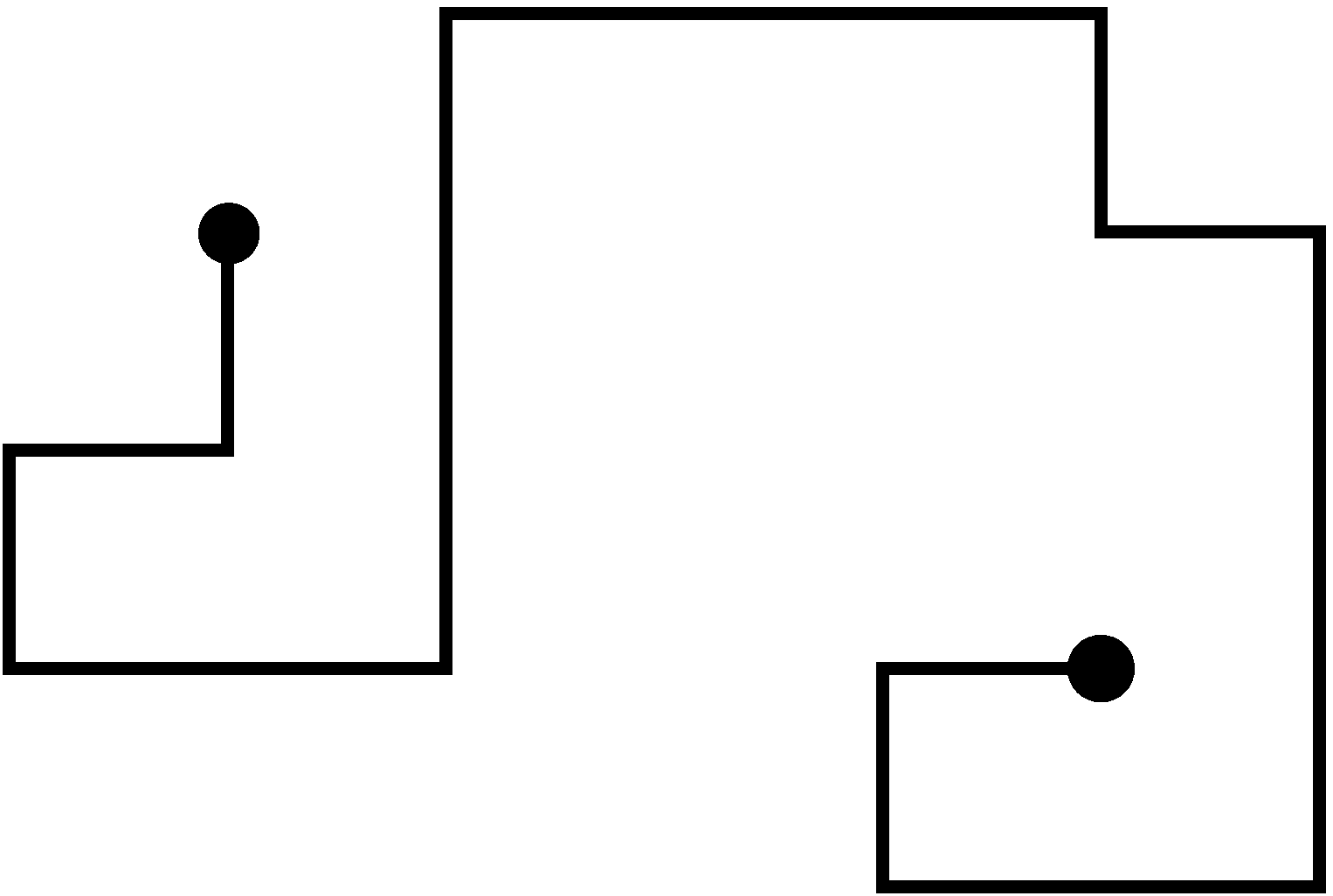}}}
\vskip .5cm \smallskip}
\caption{The derivative of the spin two-point function  in the $O(n)$
model, evaluated at $n=0$, becomes a generating function for the
self-avoiding
walks that connect two points.}
\end{figure}

\subsection{Off-critical $N=2$ supersymmetry}

\index{Supersymmetry!breaking}
The first, and most obvious property guaranteed by (unbroken)
supersymmetry in
a field
theory is that the ground state energy will be exactly zero.  This
does not
immediately
imply that the bulk free energy of the lattice model is zero since
the
partition function,
${\cal Z}$, is ambiguous up to non-universal factors of the form
$e^{fN}$,
where $N$ is
the number of ``sites'' of the model.   Equivalently, one can always
multiply
all of the
\BW by some overall function that is analytic and nowhere vanishing
in the
regime
of interest.  This will produce the foregoing ambiguity in the
partition
function.  Thus,
in a supersymmetric lattice model we should expect that the free
energy will
vanish
up to logarithms of such analytic functions.

As we remarked in the introduction, the percolation problem has $N=2$
supersymmetry in
the continuum  limit only at its critical point. To have unbroken
supersymmetry
in the
scaling region it is necessary that the trivial sector remains
trivial in the
perturbation,
and moreover that the free energy per vertex continues to be  sector
independent.
If this is so, then the free energy cannot have any singular part.
Recall that in general the free energy has two parts:
\begin{equation}
f~=~f_{\rm reg}~+~f_{\rm sing} \ ,
\end{equation}
where $f_{\rm reg}$ is analytic in the variable measuring the
distance away
from
criticality; and
\begin{equation}
f_{\rm sing}~=~A_{\pm} ~\xi^{-2} \ ,
\end{equation}
where $A_{\pm}$ are amplitudes above and below the critical point,
and
$\xi$ is the correlation length. The singular part of the free
energy, $f_{\rm
sing}$,
can be identified  with $E_0/L$ where $E_0$ is the ground state
energy of the
quantum theory. If supersymmetry
is unbroken then one has $E_0=0$, and therefore $f_{\rm sing}=0$.

This lattice criterion can immediately be used to claim that for the
polymer
problem the flow to the dilute region (where
$\beta<\beta_c$ in (\ref{cor})) does not break supersymmetry.  On the
other
hand the flow to the dense  region $\beta>\beta_c$ does break
supersymmetry.
This is
because for bigger weights, the walks fill a finite fraction of the
available
space (figure 6) \cite{BDHS}, so as long as one loop is allowed
there is a
non-trivial free energy: ${\cal Z}_R=1$ but ${\cal Z}\approx e^{fTL}$
otherwise.
Supersymmetry turns out to be broken spontaneously, a fact that is
possible
because of the non-unitarity of the perturbed polymer problem.  We
will discuss
this more in a later section.

\begin{figure}[htb]
\epsfxsize = 8cm
\vbox{\vskip .8cm\hbox{\centerline{\epsffile{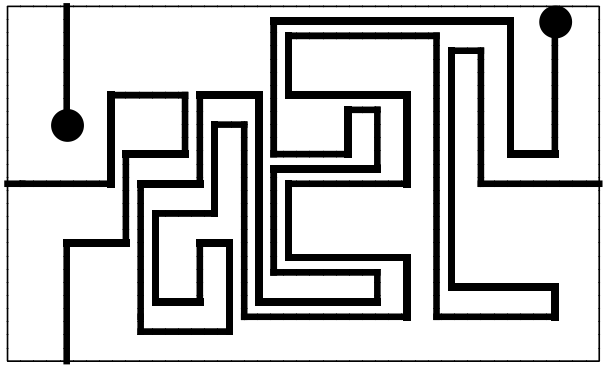}}}
\vskip .5cm \smallskip}
\caption{If the variable $\beta$ of the generating functions is real
and
bigger than the radius of convergence $\beta_c$ in infinite volume,
the
self-avoiding walks
start to fill the system and properties depend  very much on the
boundary
 conditions. The free extremities of the walks repel one another
algebraically.}
\end{figure}

If one starts with the six vertex model at criticality, then the
eight vertex model is the obvious off-critical generalization.
In the regime where the elliptic nome is positive, and for
$q=e^{i\pi/3}$, the supersymmetry is not spontaneosly broken.
(This will be established as a consequence of the results in the next
section.)
Indeed, it has been known for some time that the
singular part of the free energy vanishes for the eight-vertex model
for this value of $q$ \cite{Baxter,ABF}.
\index{Landau-Ginzburg}
\index{Eight-vertex model}
In the continnum limit, the eight-vertex model can be thought of as
a theory with \LG potential $W = X^3 - \lambda X$. In the other
regimes, where the theory is non-unitary, we expect that
supersymmetry
will be broken.

We believe that the foregoing is a generic property of $N=2$ lattice
models:
there will be an off-critical regime in which supersymmetry is
unbroken,
and a non-unitary regime where its spontaneously broken.

\subsection{Summary}

At the end of this long, and perhaps somewhat confusing, section we
would like
to
recall the properties that we expect of an $N=2$ lattice model and
summarize
them
in the form of a strategy that can be used in seeking out such
lattice models.

For any Lie algebra, spin  and quantum group parameter $q$ there is a
natural
lattice
algebra with which solutions of the Yang-Baxter equations are built.
This
algebra has
several types of representations, including a RSOS one, a vertex one,
and some
others
of mixed type \cite{kyotoa}.
Choose the value of $q$ such that the RSOS representation is trivial.
One obtains then a candidate for a topological sector.  Study then
the complete
model obtained from the vertex representation or maybe the mixed one.
Compute
in particular its torus partition function and compare it with known
$N=2$
ones.
To investigate unbroken supersymmetry away from the critical point,
study
integrable
deformations of the critical Boltzmann weights, and (i) check that
the
free energy has no singular part  (ii) compute the scaling dimension
and $U(1)$
charge of the lattice operators corresponding to the integrable
perturbation
and verify
that the continuum limit of these operators preserves the
supersymmetry.

\section{The coset construction of $N=2$ lattice models}

Thus far, we have considered only the simplest of the $N=2$ lattice
models:
those whose qunatum group structure is $U_q(SU(2))$ and whose Hilbert
space
forms a representation of the Temperly-Lieb algebra.  The critical,
continuum
limits of these models are the $N=2$ superconformal minimal models.
To go
beyond this and obtain theories from models with larger quantum group
symmetries, and whose continuum limits are the $N=2$ superconformal
coset models, one first finds a formulation of the coset models in
which the
topological sector has a simple lattice formulation.

Consider the $N=2$ superconformal coset model of the form
\cite{KazSuz}:
\begin{equation}
{\cal M}_k(G;H) ~\equiv~{{G_k \times SO_1(dim(G/H))} \over H} \ ,
\label{hssmodel}
\end{equation}
where $G$ is simply laced and $G/H$ is a hermitian symmetric space.
It
was observed in \cite{NWtopG} that there is a natural perturbation
of this model leading to an $N=2$ supersymmetric quantum integrable
field theory, ${\cal M}_k^* (G;H)$.  Moreover, the topological
subsector
of this quantum integrable model can be identified with the
topological
coset conformal field theory:
\begin{equation}
{{G_k \times G_0} \over G_{k+0}} \ .
\end{equation}
In particular, the correlators of the topological sector of
${\cal M}_k^* (G;H)$ are all constant, and can be written in terms
of the structure constants of the fusion algebra of $G_k$.
Since lattice analogues of the $G_k \times G_\ell/G_{k+\ell}$
are well known, the basic approach should now be evident.  To see
more
precisely how the N=2 lattice model is constructed we need to
elaborate
some of the details of the superconformal model.
\index{Kazama Suzuki models}
\index{Twisted algebra}

\subsection{A Coulomb Gas Formulation}

\index{Coulomb gas}
For simplicity, take the level of $G$ in~(\ref{hssmodel}) to be
one ({\it i.e.} $k=1$)\footnote{The field theories with higher
levels,
$k$, can be obtained by including generalized parafermions, and
the lattice models with higher values of $k$ can be obtained
by fusion.}.   Let $r$ be the rank of $G$.
Since $G$ and $H$ have the same rank, the representations of the
current  algebra of $G_1$ are finitely decomposable as
representations
of the current algebra of $H_1$.
Because $G/H$ is a symmetric space, the representations of the
current algebra of $SO_1(dim(G/H))$ are finitely decomposable into
representations of the current algebra of $H_{g-h}$, where $g$ and
$h$
are the dual Coxeter numbers of $G$ and $H$ respectively.  It follows
that ${\cal M}_k (G;H)$ can be thought of as a coset model of the
form:
\begin{equation}
H_1 \times H_{g-h}/H_{g-h+1} \ ,
\label{Hcoset}
\end{equation}
but with a special choice of modular
invariant.   Because of this equivalence, one can find a Coulomb gas
formulation that directly generalizes the $SU(2)$ Coulomb gas
formulation of subsection 2.3.2.  For $SU(2)$, this was a simple
Gaussian
model, here one gets a Gaussian model with $r$ free bosons
compactified
on a scaled version of the weight lattice of $G$.  \index{Gaussian
model}
The model has
a quantum group symmetry of $U_q(H_0)$ where $H_0$ is the semi-simple
factor of $H$.  (For hermitian symmetric spaces $G/H$, the group $H$
has
the form $H=H_0 \times U(1)$, where $H_0$ is semi-simple.)  The
quantum group parameter, $q$, is the one appropriate to the
denominator
factor of the coset (\ref{Hcoset}), that is, one takes
\begin{equation}
q ~=~ e^{i \pi \over (g-h+1) + h} ~=~ e^{i \pi \over g+1} \ .
\label{qvalue}
\end{equation}
This choice fixes the radius of comapctification of the
Gaussian model to the ``supersymmetric radius.''

To reduce the
Gaussian model to the requisite coset model one must perform the
BRST reduction of the field theory, which, on the lattice, is the
quantum group truncation with respect to $U_q(H_0)$.  The model knows
it origins as ${\cal M}_k (G;H)$ essentially because the Gaussian
model
is compactified upon the scaled weight lattice of $G$.  The field
theory
also contains two operators, ${\cal X}^+$ and ${\cal X}^-$, that in
the
Coulomb gas formulation extend the generators of $U_q(H_0)$ to
a twisted form of the affine quantum group $U_q(\widehat G)$.  In the
$N=2$ superconformal model, these operators are hermitian conjugates
of
each other, they are relevant, and together provide a perturbation
that
yields a unitary $N=2$ supersymmetric quantum integrable field theory
\cite{FLMW,NWtopG}.  This is directly parallel to the situation for
non-supersymmetric quantum integrable models obtained by conformal
perturbation theory: the perturbation is usually one that leads to
an affine extension of any underlying quantum group structure.  Here,
however, one needs two perturbing operators to make the quantum
integrable
model, one of which extend $U_q(H_0)$ to $U_q(G)$ and the other
extends this
to $U_q(\widehat G)$.

\index{Topological theories}
The topological twist of any $N=2$ supersymmetric field theory can be
implemented by first replacing the energy momentum tensor by
\cite{WitTop,TESKY}:
\begin{equation}
\label{toptwist}
T^{\rm top}_{\mu \nu} ~=~ T^{N=2}_{\mu \nu} ~+~ \half \epsilon_{\mu
\rho}
{}~g^{\rho \sigma} \partial_\sigma J_\nu \ ,
\end{equation}
where $J_\nu$ is the conserved $U(1)$ charge of the $N=2$
supersymmetric
theory.  With this energy momentum tensor, two of the supercharges
become dimension zero charges and can be used as BRST charges to
truncate the original Hilbert space down to the topological Hilbert
space.

The topological twist shifts the dimension of the operators
${\cal X}^+$ and ${\cal X}^-$ so that they provide exactly the
correct charges to extend $U_q(H_0)$ to $U_q(\widehat G)$.  The
topological energy mometum tensor thus commutes with $U_q(G)$.
The type ${\cal I\!I}$ representations of $U_q(G)$ at the value of
$q$ that we
have chosen in (\ref{qvalue}) are all trivial, and the BRST
reduction,
or quantum group truncation, will result in a rigid topological
model.

In the foregoing discussion we used a minor sleight of hand that we
wish
to bring into the open.  In the continuum field theory formulation of
a
coset model there are always two quantum groups, one associated with
a
numerator  factor and the other associated with the denominator
factor.
Thus in (\ref{Hcoset}) there are two quantum groups, the one that we
have
been discussing with $q$ given by (\ref{qvalue}), and one associated
with the numerator factor of $H_{g-h}$ with  $q = e^{i \pi/g}$.  Two
supercharges can be added to the latter quantum group extending it to
a twisted form of $U_q(\widehat G)$, while the extension of the
denominator
quantum group is accomplished by the relevant perturbing operators
${\cal X}^+$ and ${\cal X}^-$ as described above.  To define the
topological
theory one is supposed to use one of the supercharges as a BRST
charge, and
this implicitly suggests that one gets the topological theory using
the
numerator copy of $U_q(G)$.  However it should be remembered that the
BRST
reduction of a Coulomb gas formulation of a coset model can be done
using either one of the two quantum groups, and we have used, and
will need
to use, the denominator copy of the quantum group.

\subsection{Formulating the $N=2$ Lattice Models}

\index{Supersymmetry!$N=2$}
To build the $N=2$ lattice models one simply reverses the foregoing
course.
One starts either with a vertex model built using the fundamental
representation, ${\cal V}$, of $G$, or with the $IRF$ model whose
heights
are the weights of $G$.  The ``topologically twisted'' transfer
matrix is
built from the $\check R$-matrix of $U_q(G)$.  If one performs the
quantum group truncation of the vertex model \cite{VPHS,VPasq}, or
the
restriction of the height model, using the complete $U_q(G)$ with
$q$ given by (\ref{qvalue}) the result is a rigid lattice model.  In
the
$IRF$ formulation each successive height will be the unique level one
fusion product of the previous height and the representation ${\cal
V}$.

To get the $N=2$ lattice model one must topologically untwist the
foregoing transfer matrix, breaking the $U_q(G)$ symmetry to
$U_q(H_0)$
in such a way that the two extra generators in $U_q(\widehat G)$
are given the same spin.  One then only performs the quantum group
truncation with respect to $U_q(H_0)$, or, in the $IRF$ formulation,
only restricts the heights to be affine highest weights of $H_0$ at
level $g-h+1$, leaving the $U(1)$ direction unrestricted.  This is
referred to as partial quantum group truncation, or a partial RSOS
model.  \index{RSOS models}
This is the procedure as it was originally implemented in
\cite{MNW}.  In \cite{MNW} the Coulomb gas formulation was
extensively analysed to  precisely define the topological untwisting
of the transfer matrix, and to provide further evidence that the
result was indeed an $N=2$ lattice model.  In this review we will
use more recent results to define the off-critical \BW in the
$IRF$ formulation of these models with $G=SU(N)$. It turns out that
in the off-critical formulation the ``untwisting'' is easy to
describe
and is rather intuitive.  We begin by reminding the reader about the
construction of RSOS models based on the weight lattice of $SU(N)$.

Introduce an orthonormal basis, $f_j$, $j=1,\dots,N$, in $\IR^N$ and
define vectors $e_j$ by $e_j = f_j -{1\over N}( f_1 + \dots +
f_N)$.  The vectors $e_j$ can be thought of as weights of the
fundamental of $SU(N)$.  Consider the oriented lattice shown in
figure 8. To each vertex assign a height of the form:
\begin{equation}
\Lambda ~=~  \Lambda_0 + \sum_{j = 1}^{N-1} ~ n_j e_j \ ,
\label{heights}
\end{equation}
where $n_j \in \ZZ$ and $\Lambda_0$ is an as yet arbitrary vector.
This ``initial vector,'' $\Lambda_0$,
will play a major role in the forthcoming discussion.

\begin{figure}[htb]
\epsfxsize = 8cm
\vbox{\vskip .8cm\hbox{\centerline{\epsffile{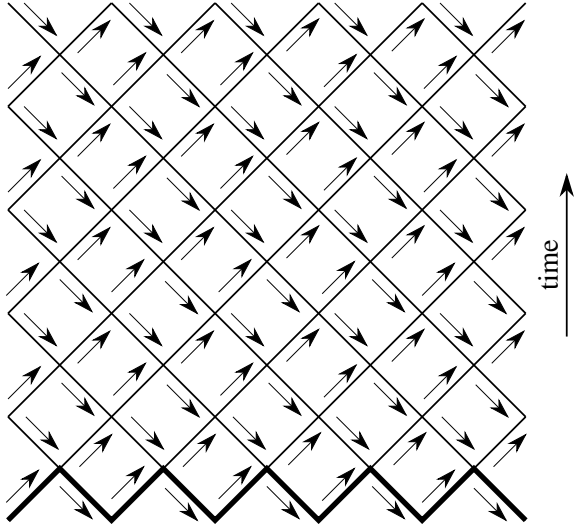}}}
\vskip .5cm \smallskip}
\caption{A section of the lattice upon which the
model is defined.  The bold zig-zag is the initial time slice, and
the
arrow indicate orientations of edges.}
\end{figure}

If $\Lambda$ is a height at the beginning of an oriented edge and
$\Lambda^\prime$ is the height at the other end of the oriented edge,
then we require that $\Lambda^\prime - \Lambda = e_j$, for some $j$.
For the typical plaquette shown in figure 9, the evolution is defined
by the \BW $w(\Lambda, \Lambda + e_i, \Lambda + e_j + e_j,
\Lambda + e_k | u)$.  Introduce the shorthand notation:
\begin{eqnarray}
[\nu] & ~\equiv~ & \vartheta_1(\gamma \nu | \tau ) \\
 & ~\equiv~ & 2 p^{1 \over 8} sin(\pi \gamma \nu)~
\prod_{n=1}^\infty
(1 - p^n)(1 - e^{2 \pi i \gamma \nu} p^n) (1-e^{-2 \pi i \gamma \nu}
p^n) \ ,
\label{sqbrack}
\end{eqnarray}
where $q = e^{i \pi \gamma}$ is the quantum group parameter and
$p \equiv e^{2 \pi i \tau}$ is the elliptic nome.
Let $\rho$ denote the Weyl vector of $SU(N)$:
\begin{equation}
\rho ~=~ \half \Big( (N-1) e_1 + (N-3) e_2 + \dots \dots
- (N-1) e_N \Big) \ .
\end{equation}
The non-vanishing \BW are as follows \cite{kyotoa}:
\begin{eqnarray*}
w(\Lambda, \Lambda + e_i, \Lambda + 2 e_i; \Lambda + e_i | u)
&  ~=~   & {[u+1] \over  [1]}  \\
w(\Lambda, \Lambda + e_i, \Lambda + e_i + e_j; \Lambda + e_i | u)
&  ~=~   & {[(\Lambda + \rho)\cdot(e_i - e_j) ~-~ u] \over
[(\Lambda + \rho)\cdot(e_i - e_j)]}  \\
w(\Lambda, \Lambda + e_i, \Lambda + e_i + e_j; \Lambda + e_j | u)
&  ~=~   &
\end{eqnarray*}
\begin{equation}
\qquad \qquad\qquad \quad
{[u] \over  [1]} ~  \left( {[(\Lambda + \rho)\cdot(e_i - e_j) + 1]
[(\Lambda + \rho)\cdot(e_i - e_j) - 1 ] \over
[(\Lambda + \rho)\cdot(e_i - e_j)]^2} \right)^\half  \ ,
\label{BWsSUN}
\end{equation}
where $i \ne j$.

\begin{figure}[htb]
\epsfxsize = 8cm
\vbox{\vskip .8cm\hbox{\centerline{\epsffile{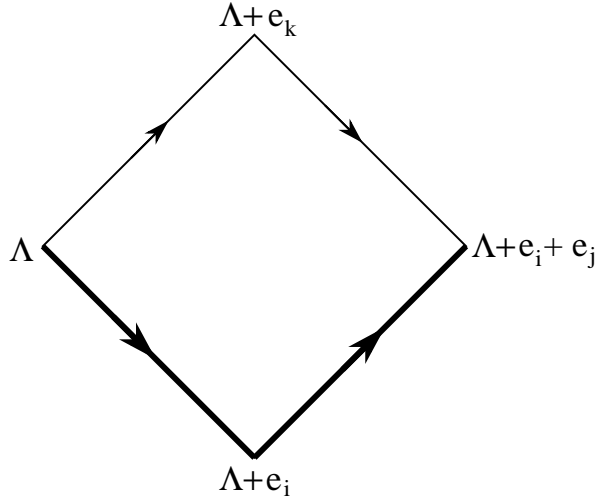}}}
\vskip .5cm \smallskip}
\caption{In the IRF formulation the heights, $\Lambda$, are
associated to vertices as shown. }
\end{figure}

\index{RSOS models}
To obtain the restricted (RSOS) models corresponding to
$G_\ell \times G_1/G_{\ell + 1}$, one takes $\Lambda_0 = 0 $,
$\gamma = 1/(\ell + g +1)$ and restricts the heights to
the fundamental affine Weyl chamber of $G_{\ell+1}$.  That is, the
heights
must be affine highest weights of $G_{\ell + 1}$.  If one were to
make this
choice of $\Lambda_0$ and $\gamma$, but not make this restriction on
the
heights, then the \BW would be singular for certain combinations of
heights.
It is elementary to verify that the transfer matrix arising from
the \BW in (\ref{BWsSUN}) preserves this restriction. The
topological, or rigid
model corresponds to setting $\ell=0$.

The lattice analogues \cite{MNW} of the $N=2$ superconformal
grassmannian models:
\begin{equation}
{\cal G}_{1,m,n} ~\equiv~ {{SU_1(m+n) \times SO_1(2mn)}
\over {SU_{n+1}(m) \times SU_{m+1}(n) \times U(1) }} \ ,
\label{grassmannian}
\end{equation}
are constructed by once again taking the heights as in
(\ref{heights})
(with $N=n+m$).   However one now performs the partial restriction by
requiring that the heights be affine highest weights of both of the
subgroups $SU_{n+1}(m)$  and $SU_{m+1}(n)$  of $SU(m+n)$, but one
does
not restrict the heights in the $U(1)$ direction.  The value of $q$
is
taken to be the ``topological value'' (\ref{qvalue}).
The topological untwisting is accomplished by choosing
\begin{equation}
\Lambda_0 ~=~ {1 \over \gamma g} (\rho_G - \rho_H)~\tau \ ,
\label{lambdazdefn}
\end{equation}
where $\rho_G$ and $\rho_H$ are the Weyl vectors of $G$ and $H$.  For
the
grassmannian model (\ref{grassmannian}), this is simply
\begin{equation}
\Lambda_0 ~=~ {m+n+1 \over 2(m+n)}[n(e_1 + \dots + e_m)
- m(e_{m+1} + \dots e_{m+n})] \tau \ .
\label{lmbdgrass}
\end{equation}
There are several things to note about this choice.
First, the proof in \cite{kyotoa,kyotob} that the \BW
(\ref{BWsSUN}) satisfy the star-triangle relations works for a
general
value of $\Lambda_0$ so this model is indeed still exactly solvable.
Secondly, $\Lambda_0$ is orthogonal to the $SU(m)$ and $SU(n)$
directions
and so these sectors of the model behave as if $\Lambda_0$ were zero.
This means that the transfer matrix preserves the partial restriction
on
the subgroup $SU(m) \times SU(n)$.  Hence this
model is an $IRF$ formulation of a particular form of (\ref{Hcoset}).
Because the $U(1)$ direction is unrestricted, there is, in principle,
a danger that the \BW could be singular for some configuartions.
In addition, in the physical model, the vector (\ref{lambdazdefn})
is purely imaginary, and so one might be concerned that
the \BW may cease to be real in the physical regime.
As we will see momentarily the Bolzmann weights are still real and
non-singular.

It is convenient to think of the foregoing models from slightly
different
perspective: the heights can be taken on the partially
restricted weight  lattice of $G = SU(m+n)$ with  $\Lambda_0 =0$, but
the
\BW must now  incorporate the shift (\ref{lambdazdefn}). We will
henceforth
adopt this perspective.  The shift by $\Lambda_0$ does not modify the
\BW in
the $H_0 = SU(m) \times SU(n)$ direction, but the other \BW are,
significantly modified.  The effect of this shift is to replace
certain strategic $\vartheta_1$'s by $\vartheta_4$'s, and this neatly
removes all the potential singularities in the \BW owing to the
unrestricted $U(1)$. In order to describe the results explicitly let
$[\nu]$
be given by (\ref{sqbrack}) and let $\{\nu\}$ is defined by:
\begin{eqnarray}
\{\nu\} & ~\equiv~ &\vartheta_4 (\gamma \nu | \tau ) \\
 &~\equiv~ & \prod_{p=1}^\infty (1 - p^n)(1 - e^{2 \pi i \gamma \nu}
p^{n-\half} )  ~(1-e^{-2 \pi i \gamma \nu} p^{n - \half}) \ .
\label{curlybrack}
\end{eqnarray}
Set the parameter $\gamma$ to ${1 \over m+n+1}$, and take the vector
$\Lambda_0$ to be zero.  The non-vanishing \BW are then as follows:

\index{Boltzmann weights}
\index{Yang-Baxter equation}
\index{Elliptic functions}
\noindent If $1 \le i,j \le m$ {\it or} $m+1 \le i,j \le m+n$ then,
for
$i \ne j$, one has:
\begin{eqnarray*}
w(\Lambda, \Lambda + e_i, \Lambda + 2  e_i; \Lambda + e_i | u) &~=~&
{[u+1] \over  [1]} \\
w(\Lambda, \Lambda + e_i, \Lambda + e_i +  e_j; \Lambda + e_i | u)
&~=~&
{[(\Lambda + \rho)\cdot(e_i - e_j) ~-~ u] \over
[(\Lambda + \rho)\cdot(e_i - e_j)]} \\
w(\Lambda, \Lambda + e_i, \Lambda + e_i +  e_j; \Lambda + e_j | u)
&~=~&
\end{eqnarray*}
\begin{equation}
\qquad \qquad \qquad \quad
{[u] \over  [1]} ~  \left( {[(\Lambda + \rho)\cdot(e_i - e_j) + 1]
[(\Lambda + \rho)\cdot(e_i - e_j) - 1 ] \over
[(\Lambda + \rho)\cdot(e_i - e_j)]^2} \right)^\half \ .
\label{ellbwsa}
\end{equation}
\noindent If $1 \le i \le m$ {\it or} $m+1 \le j \le m+n$, or
vice-versa,
then one has
\begin{eqnarray*}
w(\Lambda, \Lambda + e_i, \Lambda + e_i + e_j; \Lambda + e_i | u)
&~=~&
{\{(\Lambda + \rho)\cdot(e_i - e_j) ~-~ u\} \over
\{(\Lambda + \rho)\cdot(e_i - e_j)\}} \\
w(\Lambda, \Lambda + e_i, \Lambda + e_i + e_j;  \Lambda + e_j | u)
&~=~ &
\end{eqnarray*}
\begin{equation}
\qquad \qquad \qquad \quad
{[u] \over  [1]} ~  \left( {\{(\Lambda + \rho)\cdot(e_i - e_j) + 1\}
\{(\Lambda + \rho)\cdot(e_i - e_j) - 1 \} \over
\{(\Lambda + \rho)\cdot(e_i - e_j)\}^2} \right)^\half \ .
\label{ellbwsb}
\end{equation}

{}From the Coulomb gas analysis of \cite{MNW} and as confirmed by
the Bethe Ansatz computations in \cite{ZMa}, the foregoing height
model
at criticality yields the Neveu-Schwarz
sector of the $N=2$ superconformal model.  The Ramond sector can be
obtained by
a uniform spectral flow in the $U(1)$ direction.  This is easily
implemented:
one shifts all the lattice height appropriately, or equivalently one
takes
$\tau \to \tau + 1$ in (\ref{lambdazdefn}).

\subsection{Some simple properties of these models}

As has already been mentionned, a signal that a model is
supersymmetric
is that the free energy per lattice site is analytic when the
supersymmetry
is unbroken.   For the models discussed in the last subsection, this
computation is very revealing in that it also exhibits and utilizes
the
connection between the supersymmetric model and its topological
sector.
We will begin, however, by illustrating the foregoing construction
for
$G = SU(2)$, and relate it to the constructions of section 2.

\subsubsection{A simple example}

\index{Eight-vertex model}
For $G=SU(2)$ the foregoing lattice model
construction collapses to what is basically the $IRF$ version of the
eight-vertex model \cite{Baxter,ABF}.   The weight vectors
$e_1$ and $e_2$ of the foregoing subsection satisfy $e_1 = -e_2$, and
the Weyl
vector is given by $\rho = e_1$.  The unrestricted heights lie on
the weight lattice of $SU(2)$, which we will parametrize by an
integer,
$\ell$, where $\Lambda = \Lambda_0 + (\ell-1) e_1$.  The integer,
$(\ell-1)$,
is equal to twice the spin of the corresponding $SU(2)$ weight.
These are exactly the heights discussed in section 2.3.1.
If one takes
$\Lambda_0 = 0$ and $\gamma = 1/3$ then the quantum group truncated
model
is completely rigid with $\ell$ alternating between $1$ and $2$.
There are thus two states in this model depending upon whether a
given site
has height $1$ or $2$.  This is the $SU_1(2)/SU_1(2)$ lattice model.
To get
the lattice analogue of the $N=2$ supersymmetric theory,
one should only quantum group truncate with
respect to the $H$ subgroup, or more precisely, with respect to the
semi-simple part, $H_0$, of $H$.  In this instance, this means that
one performs no quantum group truncation at all, and one therefore
has a
Gaussian model.  The height, $\ell$, takes the values $0,1$ or $2$
modulo $3$,
and values of $\gamma$ and $\Lambda_0$ are:
\begin{equation}
\gamma ~=~ {1 \over 3} \qquad {\rm and } \qquad
\Lambda_0 ~=~ {3 \over 2}~e_1~\tau \ .
\end{equation}
We have thus ``unfrozen'' the heights $\ell$ of the RSOS model and
have avoided
the problem  of singular \BW through the choice of $\Lambda_0$.
As described in section 2.3.2, this value of $\gamma$ means that at
criticality and in the continuum the Gaussian
model is the $N=2$ superconformal minimal model, with
central charge, $c=1$.  The off-critical $N=2$ supersymmetic,
quantum integrable model corresponds to the most relevant chiral
primary
perturbation of the conformal model \cite{FMVW}, and may be thought
of
as sine-Gordon at the supersymmetric value of the coupling constant.
{}From
(\ref{ellbwsa}) and (\ref{ellbwsb}) the elliptic \BW are the
following:
\begin{eqnarray*}
w(\ell, \ell \pm 1, \ell \pm 2; \ell \pm 1 | u) &=&
{\vartheta_1({  1 \over 3} (u+1) | \tau) \over
\vartheta_1({ 1 \over 3} | \tau)} \qquad \qquad \qquad   \\
 w(\ell, \ell \pm 1 , \ell; \ell \pm 1 | u) &=&
{ \vartheta_4({ 1 \over 3}(\ell  \mp u) | \tau) \over
 \vartheta_4({ \ell \over 3} | \tau) } \qquad \qquad \qquad   \\
 w(\ell, \ell \pm 1 , \ell; \ell \mp 1 | u) &=&
{\vartheta_1({ u \over 3}  | \tau) \over \vartheta_1({ 1 \over 3} |
\tau) }
\left( {{ \vartheta_4({ 1 \over 3} (\ell - 1)
 | \tau)~  \vartheta_4({ 1\over 3} (\ell + 1) | \tau) } \over
 (\vartheta_4({  \ell \over 3}  | \tau))^2} \right)^\half \ .
\end{eqnarray*}

These \BW are precisely the same as those of the $A_2^{(1)}$ cyclic
solid-on-solid models described in \cite{Baxter,CycSOS}.  In this
context the labelling of the model by  $A_2^{(1)}$ refers to the fact
that,
because of the periodicity of the Boltzmann weights, the unrestricted
$U(1)$ is, in fact, cyclic and so rather than taking the heights to
be
in $\ZZ$ one can view them as living on the extended Dynkin diagram
of
$A_2$.

\subsubsection{The free energy}

\index{Free energy}
The easiest way to compute the free energy per unit volume, $\scf$,
is to use its analytic and inversion properties (see chapter 13 of
\cite{Baxter}).  In this review we will only consider regime $I\! I\!
I$ of the
$SU(N)$ model, {\it i.e.} $ 0 < p \equiv e^{2 \pi i \tau} < 1$,
$-\half N < Re(u) < 0$.  (This regime corresponds to the {\it
unitary},
quantum integrable field theory of interest -- other regimes either
have
different  conformal limits or correspond to perturbations with
imaginary
coupling.)  We will also consider the models defined by
(\ref{curlybrack}) -- (\ref{ellbwsb})
where $\gamma$ is now an arbitrary, positive, real parameter.

To compute $\scf$ it is first convenient to multiply the \BW
(\ref{curlybrack}) -- (\ref{ellbwsb}) by a factor of
\begin{equation}
e^{{i \pi \gamma^2 \over \tau} (u^2 + {2u \over N})} \ ,
\label{invfactor}
\end{equation}
and perform the modular inversion $\tau \to - 1/\tau$.  After making
a gauge
transformation one then finds that the \BW are manifestly periodic
under:
\begin{equation}
 u ~\to~ u ~+~ {2 \tau \over \gamma}  \ .
\label{periodicity}
\end{equation}
One then writes
$$
\scf ~=~ - \ log(\kappa(u)) \ ,
$$
and requires that $log(\kappa) $ be analytic in a region containing
regime
$I\! I\! I$ and also be periodic under (\ref{periodicity}).  Note
that one does
not impose the other periodicity of the theta functions ($ u \to u +
{2 \over \gamma}$) because such shifts of $u$ would go outside the
domain of  analyticity of the free energy.

\index{Corner transfer matrix}
Using the properties of the corner transfer matrix one can show that
$\kappa(u)$ must also satisfy:
\begin{equation}
\kappa(u) ~ \kappa(-u) ~=~ h(1-u) ~ h(1+u)  \ ,
\label{quadone}
\end{equation}
\begin{equation}
\kappa(\lambda + u) ~ \kappa(\lambda - u) ~=~ h(\lambda - u)
{}~ h(\lambda + u) \ ,
\label{quadtwo}
\end{equation}
where $\lambda = -N/2$ for $SU(N)$, and
\begin{equation}
h(u) ~\equiv~ {\vartheta_1 ({u \gamma \over \tau} |
{-{1 \over \tau}}) \over \vartheta_1 ({ \gamma \over \tau} |
{-{1 \over \tau}}) } \ .
\label{hdefn}
\end{equation}
The function $h(u)$ that appears on the right hand side of these
equations
depends only upon the inversion relations of the elliptic Boltzmann
weights.
These equations, along with analyticity and periodicity, determine
$log(\kappa(u))$ completely.  One simply writes $log(\kappa(u))$ as a
general
Fourier series in $e^{i \pi \gamma u/\tau}$ and uses (\ref{quadone})
and
(\ref{quadtwo}) to determine the coefficients.  This is a little
tedious but
it is straightforward.

At this point one should note that the foregoing equations and
constraints
{\it do not depend on the choice of}  $\Lambda_0$. This means that
the free
energy for all of of the lattice analogues of the grassmannian models
(\ref{grassmannian}) only depends upon $N = m+n$, and this free
energy is exactly that of the topological $SU_1(N) \times
SU_0(N)/SU_1(N)$
model.  This will be a general feature of these models, the free
energy
of the models will be equal to that of the rigid topological model,
and thus it must be possible to normalize the elliptic \BW
analytically
so that the free energy is zero.

The fact that the free energy is independent of $\Lambda_0$ also
means that we can use the known results for $\Lambda_0 =0$
for the models based on $SU(N)$ \cite{MPRCAT}.  One therefore has
\begin{equation}
log(\kappa(u)) ~=~ log \left( {\vartheta_1 (\gamma (u+1) |
\tau) \over \vartheta_1 (\gamma  | \tau) } \right) ~-~
\sum_{k = -\infty}^{k = \infty} ~ f(k; u, \gamma, \tau) \ ,
\label{freeen}
\end{equation}
where
\begin{equation}
f(x; u, \gamma, \tau) ~\equiv~ {{sinh\big({\pi i \over \tau}
(1 - \gamma) x \big) ~ sinh\big({2 \pi i \over \tau}  \gamma u x
\big)~
sinh\big({\pi i (N-1) \over \tau}  \gamma  x \big)} \over
{x ~ sinh\big({\pi i \over \tau} x \big)~ sinh\big({\pi i N \over
\tau}
\gamma x \big)}} \ .
\label{fdefn}
\end{equation}
This expression differs slightly from that of \cite{MPRCAT} in that
we have
subtracted the logarithm of the phase (\ref{invfactor}) so that the
result is
no longer exactly periodic under (\ref{periodicity}), but so that it
does give
the free energy for the model defined by the \BW in
(\ref{ellbwsa}) and (\ref{ellbwsb}).

To obtain the result as a function of $p = e^{2 \pi i \tau}$ one
needs to
perform the modular inversion of the second term in (\ref{freeen}).
This can
be done by Poisson resummation (see, for example, chapter 10 of
\cite{Baxter}, or appendix D of \cite{kyotoc}).  That is, one defines
\begin{equation}
\hat f (\zeta) ~\equiv ~ \int_{-\infty}^{+\infty}~
e^{2 \pi i \zeta x} ~ f(x) ~ dx
\end{equation}
and uses the equality:
\begin{equation}
\sum_{k = -\infty}^{k = \infty} ~ f(k) ~=~
\sum_{k = -\infty}^{k = \infty} ~ \hat f(k) \ .
\end{equation}
The fourier transform of (\ref{fdefn}) can be performed by closing
the contour
above or below the real axis, depending upon the sign of $\zeta$, and
then
summing the residues.  This sum over residues and the sum in
(\ref{freeen})
generates an expansion in powers of $p$.  The vanishing of
$sinh({\pi i N \over \tau} \gamma x )$ in denominator of
(\ref{freeen})
gives rise to residues that are proportional to $p^{1 \over N
\gamma}$.
This is the source of the non-analytic behaviour of the free
energy as a function of $p$.  The residues coming from the vanishing
of the
other $sinh$ function in the denominator of  (\ref{freeen}) are all
proportional to integral powers of $p$.  It is easy to extract the
leading
behaviour as $\tau \to i \infty$ from this sum over residues.  One
finds
\begin{equation}
\scf ~\sim~ {{ 4 ~sin \big({1 \over N} \big ({1 \over \gamma}
 -1 \big) \pi \big)  ~ sin \big({2 \pi u \over N } \big)~
sin \big({\pi \over N } \big)} \over
{ sin \big({\pi  \over N \gamma } )} } ~~p^{1 \over N \gamma}  \ .
\end{equation}
\index{Scaling limit}
For generic values of $\gamma$ this means that $\scf \sim
p^{1 \over N \gamma}$, while from hyperscaling one has $\scf \sim
{1 \over \xi^2}$, where $\xi$ is the correlation length.  Therefore,
we have
\begin{equation}
\tau ~\sim~ { i N \gamma \over  \pi} ~ log (\xi) \ ,
\label{tauxireln}
\end{equation}
as $\tau \to i \infty$.

One should note that for
\begin{equation}
\gamma^{-1} ~=~ N j ~+~ 1,  \qquad j =1,2,3, \dots
\label{gensusy}
\end{equation}
the numerator of (\ref{fdefn}) vanishes whenever $sinh({\pi i N \over
\tau}
\gamma x )$ vanishes.  As a result, the Poisson resummation of
(\ref{freeen})
gives rise to an expansion in integral powers of $p$.  That is, if
$\gamma$ satisfies (\ref{gensusy}) then the free energy is an
analytic function
of $p$.  The choice $\gamma^{-1} = N+1 $ corresponds to our
supersymmetric
models.

For these special values of $\gamma$ it is, of course, no longer true
that $\scf \sim  p^{1 \over N \gamma}$ as $\tau \to i \infty$.
However,
continuity in $\gamma$ means that the relation (\ref{tauxireln}) is
true
even at these special values.

It turns out that for $\gamma$ given by (\ref{gensusy}) one can
easily
express the free energy in terms of theta functions.  Here we will
give the
result for the supersymmetric model: $\gamma^{-1} = N+1 $.
The general case is similar. For $\gamma = 1/(N+1)$, the second term
in
(\ref{freeen}) can be written:
\begin{equation}
{2 \pi i \over \tau}~ u \gamma (1 - 2 \gamma) ~-~
\sum_{k = 1}^{\infty} ~ {1 \over k(1 - \tilde p ^k)} ~
(z^k - z^{-k}) ~ (w^{-k} - w^{-kN}) \ ,
\label{stdseries}
\end{equation}
where $\tilde p = e^{-2 \pi i /\tau}$, $z = e^{2 \pi i \gamma u /
\tau}$ and
$w = e^{2 \pi i \gamma  / \tau}$.  This is a standard expansion of
the
logarithm of the ratio of two theta functions\footnote{One simply
takes the
logarithm of the product formula for the theta functions, expands all
of
the $log(1 -q^n x^{\pm 1})$ terms into power series in $q^n x^{\pm
1}$, and
then  one can perform the sum over $n$ to obtain (\ref{stdseries}).}.
Finally one can perform modular inversions on the theta functions and
the
final result is
\begin{equation}
log(\kappa(u)) ~=~ log \left( {\vartheta_1 (\gamma (u-1) |
\tau ) \over \vartheta_1 (\gamma |  \tau )} \right) ~=~ log \left( {
[u - 1] \over [1] } \right) \ .
\end{equation}

Observe that, as promised, this free energy is analytic in the whole
of regime
$I\! I \! I$.  Indeed, if one multiplies all of the \BW in
(\ref{ellbwsa}) and
(\ref{ellbwsb}) by ${[1] \over [u - 1]}$ then the \BW are still
analytic
in regime $I\! I \! I$, and the free energy, $\scf$, vanishes
identically.

\subsubsection{The off-critical perturbing operators}

\index{Conformal field theory!perturbed}
\index{Supersymmetry!$N=2$}
For the critical limit of these models one can use
Coulomb gas methods to show that the continuum limit is
$N=2$ supersymmetric.  The free energy computations suggest that the
off-critical models are also $N=2$ supersymmetric in the continuum,
but
we would like to demonstrate this more directly.

{}From the analysis of the perturbations of $N=2$ superconformal
coset models,
we know that the are natural perturbations that lead to $N=2$
supersymmetric
quantum integrable models.  These operators have the form:
\begin{equation}
\psi ~\equiv~ \Gminus \widetilde \Gminus \phi \qquad {\rm and} \qquad
\tilde \psi ~\equiv~ \Gplus  \widetilde \Gplus \tilde \phi \ ,
\label{susyperts}
\end{equation}
where $\phi$ is a very particular chiral primary field, and $\tilde
\phi$
is its anti-chiral conjugate.  In particular, the fields $\psi$ and
$\tilde \psi$ have conformal  weights and $U(1)$ quantum numbers:
\begin{eqnarray}
h_\psi ~=~ \bar h_\psi ~=~ h_{\tilde \psi} ~=~ \bar h_{\tilde \psi}
&~=~& \half ~+~ {1 \over 2(N+1)} \nonumber \\
Q_\psi ~=~ \bar Q_\psi ~=~ - Q_{\tilde \psi} ~=~ - \bar Q_{\tilde
\psi}
&~=~& 1 ~-~ {1 \over (N+1)}
\label{susyqnumbs}
\end{eqnarray}
We remarked above that in the vertex form of the model at
criticality,
these operators extend $U_q(H_0)$ to $U_q(G)$.  Thus, based on
general
expectations about lattice models, one would expect these operators
to
be the ones that correspond to the elliptic ``deformation'' of the
critical
model.  We can see this connection much more explicitly as follows.

Consider the continuous family of models where the initial height
vector is
taken to be
\begin{equation}
\Lambda_0 ~=~ {2 \mu \over \gamma g} (\rho_G - \rho_H) ~ \tau \ ,
\label{newldefn}
\end{equation}
where $\mu$ is a parameter.   When $\mu =0$ we have the topological
model
and the lowest powers of $p$ in the \BW (\ref{BWsSUN})
are $p^0$ and $p^1$.  If we increase $\mu$, then the lowest powers of
$p$
change smoothly to $p^\mu$ and $p^{(1 - \mu)}$.  \index{Scaling
limit}
Now recall that in the
scaling limit, $\tau \to i \infty$, we know that $\tau$ is related to
the correlation  length by (\ref{tauxireln}), and thus  $p \sim
\xi^{-{2N \over N+1}}$.  It follows that the elliptic perturbation
must
involve two operators, and that their coupling constants must scale
as $p^\mu \sim \xi^{-{2N \mu \over N+1}}$ and $p^{(1 - \mu)} \sim
\xi^{-{2N (1- \mu) \over N+1}}$.  The corresponding operators
therefore
have conformal weights:
\begin{equation}
h_1 ~=~ \bar h_1 ~=~ 1 ~-~ {N \mu \over (N+1)}\ ;  \quad \quad
h_2 ~=~ \bar h_2 ~=~ 1 ~-~ {N(1 - \mu) \over (N+1)} \ .
\label{pertwts}
\end{equation}
When $\mu= \half$ these two operators have precisely the same
conformal
dimension and this is the conformal dimension given in
(\ref{susyqnumbs}).
If we now consider the gradual untwisting of the energy momentum
tensor of
the quantum field theory:
\begin{equation}
T (z) ~=~ T^{\rm top} (z) ~-~ \mu ~\partial J(z) \ ; \quad \quad
\widetilde T (\bar z) ~=~ \widetilde T^{\rm top} (\bar z) ~+~
\mu ~\bar \partial  \widetilde J(\bar z) \ ,
\label{graduntwist}
\end{equation}
we see that the conformal weights of the operators
(\ref{susyperts}) depend upon the parameter $\mu$ exactly
as in (\ref{pertwts}).

In this way one can not only identify the dimensions of the
perturbations
that lead to the elliptic Boltzmann weights,  but one can also
determine their
$U(1)$
charges.  Moreover, the foregoing also demonstrates that smoothly
changing the initial lattice vector, $\Lambda_0$,  causes the
dimensions of the perturbing operators to change in exactly the way
that
they should under topological untwisting.   As a result one also
obtains
further  confirmation that one has correctly determined the lattice
analogue
of topological untwisting. A more explicit relation between
$\Lambda_0$ and
perturbed
conformal theories can be obtained using the method of \cite{NRHS}.

\section{Scattering matrices}

\index{Scattering matrices}
Since we are considering critical and off-critical $N=2$ lattice
models, there
will
of course be massless and massive $S$-matrices associated with the
excitations
of these
models.  There have been quite a number of papers written on this
subject
(see, for example,
\cite{FMVW,PMMAW,PFKI,ALDNNW,WLNW,Grisaru,Hollow}), and we
will
not try perform even a partial survey.  Our purpose here is to simply
make a
few remarks about
some of the known $S$-matrices, and how their construction fits in
with the
general
strategy of the construction of $N=2$ lattice models.

It is first useful to recall some general observations about
conformal coset models and their perturbations leading to quantum
integrable
models.  In a generalized Coulomb gas description of the conformal
model, there
are always two quantum groups that usually have the same underlying
Lie
algebra, but
have different roots of unity.  One of these quantum groups is
generically
associated
with the numerator of the coset, while the other quantum group is
associated
with
the denominator.  The corresponding lattice model only exhibits one
of
these quantum group symmetries,
and it is usually the one associated with the denominator of the
coset model.
The perturbing operators that lead to quantum integrable models
usually extend
this lattice (or denominator) quantum group to a larger, affine
quantum group.
The other (numerator) quantum group usually plays a major role in the
scattering
theory of the quantum integrable model.  That is, the scattering
matrices are
usually built out of the $\check R$-matrices of this numerator
quantum group.
The corresponding affine extension of the scattering quantum group
leads to
operators
in the theory that commute with the $S$-matrix.  Such non-local
conserved
charges
have been used extensively in the construction of the $S$-matrices.
The
intuitive
reason as to why the scattering theory and lattice quantum groups are
distinct
is that the solitons of the scattering theory must be local with
respect to
the perturbing operators leading to the quantum integrable model
\cite{GFAL}.

The supersymmetric models also fit this mould.  The denominator
quantum group
is
indeed that of the lattice model, while numerator quantum group is
that of the
scattering theory.  The non-local conserved charges that extend the
scattering
quantum
group to the affine quantum group are simply the supersymmetry
generators.
Thus we
find that while almost all of our lattice models do not have explicit
supersymmetry,
the corresponding scattering theories do. In other words
supersymmetry appears
only as a
dynamical symmetry.

\subsection{Massless $S$-matrices and the six-vertex model}

\index{Massless scattering}
\index{Bethe ansatz!thermodynamic}
\index{Six-vertex model}
The six-vertex model is diagonalizable by Bethe ansatz \cite{Baxter},
and the
solutions are classified in terms of $1,2$ strings and the
anti-string $1^-$,
with
a coupling diagram as in figure 10 (see \cite{MTMS}). To study the
scattering
theory
simply observe that for a relativistic QFT the thermodynamic Bethe
ansatz
\cite{Zamo},
in addition to the pseudo-particles appearing in the solution of
the Bethe equations, involves an additional physical particle, as in
figure 10.

\begin{figure}[htb]
\epsfxsize = 6cm
\vbox{\vskip .8cm\hbox{\centerline{\epsffile{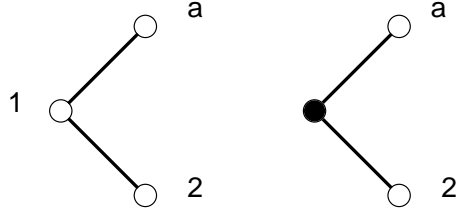}}}
\vskip .5cm \smallskip}
\caption{The first diagram represents the pseudo-particles
(solutions of the Bethe ansatz)
and their couplings at the supersymmetric
point of the lattice vertex model. Recall that for $q=e^{i\pi/t}$ one
has
strings $(1,2,\ldots,t-1)$ and the anti-string $a$.
In the continuum theory there is an additional
node for the physical particles, indicated by a heavy dot
on the second diagram. Therefore the number of Bethe ansatz
solutions
must differ by one  from  the lattice case, hence
the denominator of the quantum group parameter must also differ by
one from the
lattice case.}
\end{figure}

Therefore, if the lattice model has $q=\exp(i\pi/3)$,  the scattering
theory
has
$q'=\exp(i\pi/2)$.  From the lattice model, a massless scattering
theory
\cite{LDFLAT} can be extracted which has the following content
(see \cite{NRHS,FSrev}).  One has a pair of right moving particles
which we
parametrize as
\begin{equation}
e_R=p_R={M\over 2}e^\theta
\end{equation}
and a pair of left moving particles with
\begin{equation}
e_L=-p_L={M\over 2}e^{-\theta}
\end{equation}
where $M$ is a mass scale and $\theta$ is a rapidity. These particles
have fermion number equal to $\pm 1$ and they have factorized
scattering
with
\begin{equation}
S_{LL}=S_{RR}=Z(\theta)\left(\begin{array}{cccc}
\cosh\theta/2&0&0&0\\
0&i\sinh\theta/2&1&0\\
0&1&i\sinh\theta/2&0\\
0&0&0&\cosh\theta/2
\end{array}\right)
\end{equation}
where $Z(\theta)$ is a normalization factor ensuring unitarity and
crossing
symmetry, while the LR scattering is trivial
\begin{equation}
S_{LR}=S_{RL}=1
\end{equation}
Acting on the right  particles $u,d$ we have
\begin{eqnarray}
Q_-|u(\theta)>=\sqrt{M/2}e^{\theta/2}|d(\theta)>\\
Q_+|d(\theta)>=\sqrt{M/2}e^{\theta/2}|u(\theta)>
\end{eqnarray}
and same type of relations with left particles and $\bar{Q}^\pm$
generators.
The supersymmetry algebra is
\begin{eqnarray}
Q_+^2=Q_-^2=\bar{Q}_+^2=\bar{Q}_-^2=\{Q_+,\bar{Q}_-\}=\{Q_-,\bar{Q}_+\}=0\\
\{Q_+,\bar{Q}_+\}=\{Q_-,\bar{Q}_-\}=0\\
\{Q_+,Q_-\}=Me^\theta,\ \{\bar{Q}_+,\bar{Q}_-\}=2Me^{-\theta}
\end{eqnarray}

\subsection{Massive $S$-matrices}

\index{Landau-Ginzburg}
\index{Conformal field theory!perturbed}
There are several ways in which one can extract the $S$-matrices of
the massive
quantum integrable models that appear in the continuum limits of the
$N=2$ lattice models.   We will briefly discuss two of them here. The
first
approach is that taken in \cite{FMVW,PFKI}, and the
starting point is to use the \LG description of the model to
determine
the ground-state and soliton structure.  Since the solitons are
generically
doublets
of the superalgebra, one can then use the Bogomolnyi bounds in the
superalgebra
to determine the soliton masses.  The solitons also have fractional
fermion
numbers
that can be determined from the \LG potential.  This information,
combined with
the
fact that the $S$-matrix must commute with the supersymmetry, be
unitary,
satisfy the
Yang-Baxter equations, and obey crossing symmetry is more than enough
to
determine the
$S$-matrix completely.  This has been done for a large class of
models
\cite{PFKI}, but
there is an even larger class for $N=2$ quantum integrable models for
which the
$S$-matrices are not known.  The first reason for this is that  one
can easily
show that
in this larger class the solitons alone cannot form a kinematically
closed
scattering
theory \cite{FMVW,WLNW}.  This almost certainly means that breather
states must
be
included in the spectrum.  This does not pose, {\it a priori}, any
obstacle to
the
construction of the $S$-matrices, but as we will describe below,
there could be
some further conceptual problems to be solved if one is to determine
the
$S$-matrices
for these  more general models.

We will describe the second method of obtaining the $S$-matrices in
more detail
here
since it is rather more in the spirit of our approach to $N=2$
lattice models.
The basic idea is to once again use topological models and a
modification of
quantum group truncation.

The scattering matrices are well known for the quantum integrable
models
obtained from
the natural perturbations of the $G_k \times G_\ell/G_{k+\ell}$ coset
models
(see, for example,
\cite{TRKEM,CADBAL,THQG,TNHJVVF,PCGM,EOPW,BazRes}).
For simplicity we will only consider the models with with $G=SU(N)$
and the
level  $k$ equal to one.  The
$S$-matrices can then be obtained from RSOS restrictions of affine
Toda $S$-matrices.  Let $W_j$, $j =1,\dots,N-1$, denote the
fundamental
representations of $SU(N)$ ({\it i.e.} the anti-symmetric tensors of
rank $j$).
Each of these representations defines a multiplet of solitons of
affine
$\widehat{SU(N)}$-Toda, and these solitons can be labelled by the
weights of
the
representations.

\index{Quantum groups!representations}
Define $S^{(\widehat G)}_{ji} (\theta , q)$ to be the two-body
$S$-matrix for
the scattering of particles in $W_i$ with those in $W_j$:
\begin{equation}
S^{(\widehat G)}_{ji} (\theta, q) : \quad W_i \otimes W_j ~\to~ W_j
\otimes W_i
 \ ,
\end{equation}
where, as usual, $\theta$ is the relative rapidity of the solitons
and $q$
is the quantum group parameter.
Requiring the $S$-matrix to commute with the $U_q(\widehat{SU(N)})$
symmetry
leads to the following result:
\begin{equation}
S^{(\widehat G)}_{ji} (\theta , q) ~=~ X_{ji} (\theta )\  v_{ji}
(\theta, q) \
R_{ji} (\theta , q ) \ .
\label{sdecomp}
\end{equation}
The last term, $R_{ji} (\theta , q) $, is the standard $R$-matrix for
the
quantum group $U_q(\widehat{SU(N)})$ in the principal
gradation\footnote{The  $R$-matrices that one usually encounters in
the
literature are written in the homogeneous gradation, which is related
to
the principal gradation by a trivial automorphism, which will soon
be described.}.
The spectral parameter, $x$, and the quantum group parameter, $q$, of
the
$R$-matrix are related to the rapidity, $\theta$ and the Toda couping
constant, $\beta$, by:
\begin{equation}
q ~=~ exp\left( -{i \pi \over \beta^2} \right) \ , \quad
x~=~ exp\left( {\theta \over \gamma} \right) \qquad {\rm where}
\qquad
\gamma ~=~ {\beta^2 \over 1 - \beta^2} \ .
\end{equation}
The scalar factor, $v_{ji} (\theta , q)$, in (\ref{sdecomp}) is the
minimal
factor that makes the product $v_{ji} R_{ji}$ crossing symmetric
and unitary. This is relatively easy to compute and can be found in
\cite{CADBAL,THQG}.  The additional scalar factor $X_{ji} (\theta)$
is a CDD
factor,
and it satisfies crossing and unitarity by itself.
This factor contains all of the necessary poles for closure of the
bootstrap
with the
spectrum of soliton masses.  It turns out that this CDD factor does
not depend
upon the value of the Toda coupling and so can be evaluated by going
to the
limit
($\beta \to 1$) in which the Toda model can be related to a
Gross-Neveu model
\cite{CADBAL}.

To get the model corresponding to the perturbed coset model one must
first
incorporate the background charge, which is done in an analogous
manner to
the twist in the Markov trace of subsection 2.3.4.  Let $H$ be some
subgroup of $G=SU(N)$, where $H$ also has rank $N-1$, and let
$\rho_H$ be the
Weyl vector of $H$.  Define $S^{({\widehat G}/H)}_{ji} (\theta, q)$
by
\begin{equation}
S^{({\widehat G}/H)}_{ji} (\theta, q) ~=~ \left(x^{-\rho_H \cdot h}
\otimes x^{-\rho_H \cdot h} \right) ~ S^{({\widehat G})}_{ji} ~
\left(x^{\rho_H \cdot h} \otimes x^{\rho_H \cdot h} \right) \ ,
\label{htwist}
\end{equation}
where $h$ represents the vector of Cartan subalgebra generators of
the
quantum group.  This conjugation only affects the $R$-matrix part of
the
$S$-matrix, and for $H=G$, it converts the principal gradation to the
homogeneous gradation.

By construction, the S-matrices $S^{({\widehat G}/H)}_{ji} (\theta,
q)$
commute with the action of the finite quantum group $U_q(H)$.
These generators act in the more familiar rapidity independent
fashion.
Therefore one can use the $U_q(H)$ symmetry to restrict the
model exactly as in the lattice model.

For each weight, $\mu$, of $W_j$ we introduce a formal operator
$K^{(j)}_\mu (\theta )$.   These operators satisfy an $S$-matrix
exchange
relation:
\begin{equation}
K^{(j)}_\mu (\theta_1 ) \  K^{(i)}_\nu (\theta_2 )
{}~=~ \sum_{\mu ', \nu '}
\left( S^{({\widehat G}/H)}_{ji} (\theta, q)\right)^{\mu ' ,\nu
'}_{\mu, \nu} ~
K^{(i)}_{\nu'} (\theta_2 )\  K^{(j)}_{\mu'} (\theta_1 ) \ .
\nonumber
\end{equation}
Let ${\cal F}$ denote the multi-particle fock space generated by the
formal
action of the operators $K^{(j)}_\mu (\theta )$ on the vacuum. The
space
${\cal F}$ is an $U_q(H)$ module, and reducible (for $q$ not a root
of unity):
\begin{equation}
{\cal F} = \bigoplus_a ~ V^{( \lambda^{(H)}_a )} \ ,
\nonumber
\end{equation}
where
$ V^{( \lambda^{(H)}_a )}$ is an $U_q(H)$ module of highest weight
$ \lambda^{(H)}_a$.  Since the $K^{(j)}_\mu (\theta )$ act on ${\cal
F}$,
one can consider their reduction
\begin{equation}
K^{(n)}_{ \lambda^{(H)}_b  \lambda^{(H)}_a  } (\theta ) :
\quad  V^{( \lambda^{(H)}_a  )} \longrightarrow
V^{( \lambda^{(H)}_b )} .
\nonumber
\end{equation}
These operators satisfy the exchange relation:
\begin{equation}
K^{(j)}_{ \lambda^{(H)}_b  \lambda^{(H)}_a  } (\theta_1 ) \
K^{(i)}_{ \lambda^{(H)}_a  \lambda^{(H)}_c  } (\theta_2 ) \
{}~=~ \qquad \qquad \qquad \qquad \qquad \qquad \qquad \qquad \qquad
\nonumber
\end{equation}
\begin{equation}
\qquad \qquad \qquad \qquad
\sum_{\lambda^{(H)}_d}
\left( S^{({\widehat G}/H)}_{ji} (\theta, q)\right)^{\lambda^{(H)}_b
\lambda^{(H)}_d}_{\lambda^{(H)}_a  \lambda^{(H)}_c} ~
K^{(i)}_{ \lambda^{(H)}_b  \lambda^{(H)}_d  } (\theta_2 ) \
K^{(j)}_{ \lambda^{(H)}_d  \lambda^{(H)}_c  } (\theta_1 ) \ .
\nonumber
\end{equation}
The S-matrix for the kinks in this equation is the SOS form, and the
foregoing construction is once again the vertex/SOS correspondence.
\index{Vertex/IRF correspondence}

\index{Qunatum group truncation}
As before, the restriction amounts to taking $q$ to be a root of
unity and
imposing a limitation on the allowed highest weight labels
$\lambda^{(H)}$.
To get the $S$-matrix for the preturbed coset model
$G_1 \times H_\ell/H_{\ell + 1}$ one must take the Toda coupling so
that
$\beta^2 = (\ell + h)/(\ell + h +1)$, where $h$ is the dual Coxeter
number
of $H$.  Therefore, one has:
\begin{equation}
q ~=~ - exp(- i \pi /(\ell +h ) ) \ .
\label{funnyq}
\end{equation}
Note that this is the root of unity appropriate to the quantum group
structure of the numerator quantum group of the coset
model\footnote{In the earlier sections of this review, where we were
discussing lattice models, we took
$q = e^{i\pi/M}$, where $M$ was some positive integer.  We could
equally well
have taken $q$ to be defined by $q = - e^{-i\pi/M}$.  Here, however,
the
affine extension of the quantum group is now playing a more central
role
and this means that one has to make a specfic choice for $q$.  This
choice is,
of course, convention dependent and we are employing the conventions
of
\cite{ALDNNW,CADBAL}.}.

To get the $N=2$ supersymmetric models one chooses $H$ so that
$G/H$ is hermitian symmetric, and takes $\ell = g - h$.  Thus one has
$q = - e^{- i \pi/g}$.  If one were to take $H=G$ one would once
again
get the topological coset model, but with the foregoing choice of
$q$,
the only $U_q(G)$ representation with non-vanishing $q$-dimension is
the
singlet representation.  There are thus no solitons in the
topological model.
Once again one finds the $N=2$ model by modifying the quantum group
truncation procedure.  One can also verify that under the twisting
process
(\ref{htwist}), the two generators that extend $U_q(H)$ to
$U_q(\widehat G)$
have a rapidity dependence that makes them spin-$\half$ charges.  If
the
quantum group truncation procedure parallels that of the conformal
theory,
then these operators will become the supersymmetry generators of the
restricted model.

This process works beautifully and simply for $G = SU(N)$ and
$H = SU(N-1) \times U(1)$.  For this choice, the fundamental
reprexentations $W_j$ decompose into a direct sum of two
representations
of $U_q(SU(N-1))$.  In the RSOS truncation using $U_q(H)$ one
therefore
finds that each $W_j$ yields two solitons, and these form a
supersymmetric
doublet.  The $U(1)$ charge becomes the (fractional) fermion number,
and the
two generators that extend $U_q(H)$ to $U_q(\widehat G)$ act as the
supersymmetry
on this doublet.  We have thus incorporated the supersymmetry
directly
into the affine quantum group. The $S$-matrices for the $N=2$ model
can then be
easily extracted from the Toda $S$-matrices for $G = SU(N)$
\cite{ALDNNW}.

The surprise comes once one tries this procedure for more general
models of the
form
(\ref{grassmannian}), for example, for  $G=SU(4)$, $H=SU(2) \times
SU(2) \times
U(1)$.
The problem is that the quantum group truncation does not respect
canonical
supermultiplet structure.   In these more general models
all of the $W_j$'s give rise to RSOS solitons with
fractional fermion numbers that agree with the \LG picture.  All of
the
RSOS solitons coming from one of the $W_j$'s are, of course,
transformed into
each other by the generators of $U_q(\widehat G)$.  The  two
generators that
extend
$U_q(H)$ to $U_q(\widehat G)$ are indeed spin-$\half$ and commute
with the
$S$-matrix.  The problem is that there does not seem to be an RSOS
truncation upon which the algebra of these generators becomes that of
the standard $N=2$ superalgebra.  For example, some $W_j$'s would
give rise
to a three dimensional ``supermultiplet.'' (This happens in the
$SU(4)$ example
above for the six-dimensional representation of $SU(4)$.)

There are several possible resolutions of this issue.
It is quite conceivable that the Toda approach breaks down, or it
might be
that there is some supersymmetry anomaly.  It is also possible that
the integrable model gives rise to two $S$-matrices, one in which
the supersymmetry acts locally, and one in which it does not.
It is certainly of interest to resolve this problem.   As first
sight,
the possibility of a supersymmetry anomaly seems the least likely
explanation.
However, such an explanation is

perhaps not quite as outrageous as it sounds, after all, the
supersymmetry
would still be a symmetry of the $S$-matrix, but it would simply have
non-trivial
braiding relations with single soliton states.  Such a phenomenon is
already
present in the models that we do understand:  the fractional fermion
number
of single solitons means that the supersymmetry action on single
solitons
must involve extra phases.  It is just conceivable that the more
exotic models
exhibit a more involved (perhaps non-abelian) form of this.
At any rate, it is highly desirable to compute the $S$-matrices for
these
models.
It is also, perhaps, significant that these ``exotic'' models are
also
precisely the ones for which the $S$-matrices are not yet known
because the
solitons cannot, by themselves, form a closed scattering theory.

\section{Other issues}

\subsection{Spontaneous breaking of $N=2$ supersymmetry}

\index{Supersymmetry!breaking}
In lattice models there is a natural direction of perturbation that
does
not break supersymmetry explicitly, but does so spontaneously.
This is made possible
by the non-unitarity of the perturbation. For example, consider the
polymer
problem discussed earlier.  \index{Landau-Ginzburg}
The critical \LG potential in this theory is
$W = X^3$, and the perturbation that changes the weight $\beta$
of monomers in the generating functions corresponds exactly to adding
a multiple of $X$ to this potential.
However if one writes out the $F$-term term in the supersymmetric
action,
and properly incorporates the coupling constant, one obtains:
\begin{equation}
S_F ~=~{\rm const.} \left( \int d^2 z d^2 \theta \left( X^3
+(\beta_c-\beta)^{1/2}~X
\right) ~+~  \right. \qquad \qquad \qquad \qquad
\nonumber
\end{equation}
\begin{equation}
\qquad \qquad \qquad \qquad \qquad \qquad \qquad
\int d^2 z d^2 \bar \theta \left( \bar X^3 +(\beta_c-\beta)^{1/2}
{}~\bar X \right) \ .
\end{equation}
If $\beta<\beta_c$ one gets the usual physics of the off-critical
quantum integrable model with superpotential $W=X^3+X$.
On the other hand, if $\beta>\beta_c$ one gets very different physics
because
the perturbation is purely imaginary, and the model is non-unitary.
This is easily analyzed qualitatively. For
 $\beta\leq\beta_c$ polymers are ``very tiny'' (their fractal
dimension is less
than
two) so the free energy calculated in the Ramond sector, where no
polymer at
all
is allowed, is the same as the free energy calculated in any other
sector,
where some non-contractible polymers are allowed.
For $\beta>\beta_c$ the partition function of the Ramond sector is
still
${\cal Z}_R=1$ but as soon as a polymer is allowed it fills most of
the
available space and ${\cal Z}$ grows exponentially with the size of
the system.
This means that there are infinitely many level crossings
between the ultra-violet and infra-red fixed points in the Ramond
sector,
and although the states of vanishing energy always stay there, they
are
infinitely
far from the true ground state of negative energy in the infra-red.
In the infra-red the bosonic degrees of freedom
of the theory have become massive and the fermionic ones are a simple
(topologically twisted) Dirac fermion,
describing the physics of ``dense polymers''.

\index{Painleve}
Interestingly, the position of the
level crossings can be related to the poles of the Painleve III
differential
equation \cite{FSZ}. Consider the system on a cylinder of radius $R$
and length
$T$.  The
renormalization group variable $z$ is proportional to
$(\beta_c-\beta)R^{4/3}$.  Recall that the generalized supersymmetry
index is
defined by \cite{CFIV}:
\begin{equation}
Q ~=~ i{R \over T} ~tr\left( e^{-RH}~ F(-)^F~ \right)
\label{newindex}
\end{equation}
where, from the point of view of the hamiltonian evolution,
the ``time'' is in the $R$ direction now.  This index can be written:
\begin{equation}
Q={z\over 2}{du\over dz}
\end{equation}
where $u$ is the solution of the Painleve III equation:
\begin{equation}
{d^2u\over dz^2}+{1\over z}{du\over dz}={9\over 16\sqrt{-z}}\cosh u
\end{equation}
with specific asymptotic behaviour \cite{CFIV}.
One can show that the level crossings occur at the poles of $Q$ on
the negative
$z$ axis. One can
also show that these poles are simple, and that there are an infinite
number
of them \cite{FSZ}.

\subsection{Ultra-high temperature limits: the non-interacting models
revisited}

\index{Bloch walls}
The trivial non-interacting spin model of section 2 has even more to
it than
was
made evident there. Recall that the spins sit on the edges
of the original square lattice (figure 1).
Suppose that we now draw, on the diagonal lattice of figure 3,

occupied edges between spins of {\it opposite} value.

These edges correspond to a
 Bloch wall, and the loops created by them
can traverse each edge of the lattice at most once. (See figure 11.)
These are the usual
contours of the low temperature Ising model, but considered in the
high-temperature phase (actually, in the limit of infinite
temperature).
We now consider the geometrical
properties of these loops. First it should be clear from preceding
discussion
that they are identical to the properties of the boundaries of
percolation
clusters (or hulls). The most interesting quantity is the fractal
dimension
of these loops, which is related to the algebraic decay of the
probability
that two edges belong to the same loop.  That is, if the latter
decays as
$r^{-2x}$ then the fractal dimension of the loop is $D_f={ 2-x}$. One
can
show that the exponent is given by $x=2h$, where $h$ the conformal
weight,
in the twisted theory,  of the Ramond ground state with the lowest
charge. That
is:
\begin{equation}
h ~=~ {1\over 24} - {1 \over 2} \left({-{1\over 6}}\right)  ~=~ {1
\over 8}
\end{equation}
and thus $D_f={7\over 4}$. This is midway between the fractal
dimension
for brownian motion ($D_f = 2$) and the fractal dimension of
self-avoiding
random walks ($D_f= {4 \over 3}$).
It is interesting to observe that,
from the point of view of the low-temperature expansion for the Ising
model,
another geometrical quantity is very natural: the probability that
two edges
are extremities
of  the  same open line.  This corresponds to the two-point
correlation
function of disorder operators. This function
goes to a constant at large distance, and so the conformal weight is
zero.
This high-temperature limit is therefore rather interesting because
it rather naturally leads to the existence of a non-trivial operator
of
vanishing
dimension in a ``lattice topological sector.''
\index{Fractal}

\begin{figure}[htb]
\epsfxsize = 6cm
\vbox{\vskip .8cm\hbox{\centerline{\epsffile{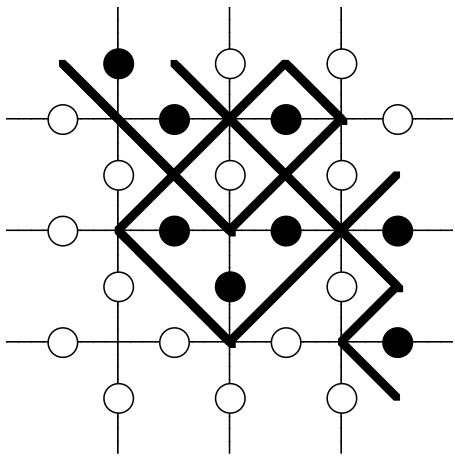}}}
\vskip .5cm \smallskip}
\caption{The Bloch walls for Ising model at high temperature.  Filled
dots
represent states of spin $+1$ and empty dots represent states of spin
$-1$.
The walls separate domains of opposite spin.}
\end{figure}

Now we have a natural direction to study: the properties of Bloch
walls
in high temperature phases of spin models. The topological sector
corresponds to the original spin variables. The non-topological one
to the
geometry
of the Bloch walls.  As another example consider the three state
Potts model.
There
are three states per vertex, represented by empty dot, full dot or
square in
figure 12.
The Bloch walls carry one arrow if, when one faces along the arrow,
there is
either
(i) an empty dot on the right and a full dot on the left, or
(ii) there is a square on the right and an empty dot on the left.
The Bloch walls carry two arrows if there is a square
is on the right and a full dot on the left of the arrows.   With
these rules there is obviously current conservation.

\begin{figure}[htb]
\epsfxsize = 6cm
\vbox{\vskip .8cm\hbox{\centerline{\epsffile{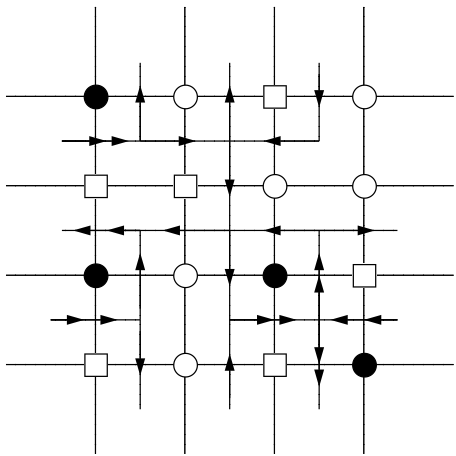}}}
\vskip .5cm \smallskip}
\caption{The Bloch walls for the three-state Potts model at high
temperature.
By
convention the full dot, empty dot and square correspond to
$\sigma=1,2,3$
respectively.
The walls
carry one or two arrows depending on the difference of spins between
both
sides. By convention the highest spin is to the right of the
arrow(s).}
\end{figure}

By analogy with the foregoing Ising model, there are two natural
geometrical
 correlation functions
that are constant at large distance: the probabilities that two edges
are
extremities of the same open
line carrying either one or two arrows.  Similarly there are two
natural
correlation functions that decay algebraically: the probabilities
that two edges belong to the same loop carrying one or two arrows. In
this
example there is charge conjugation symmetry that in fact relates
simple and
double arrows so these pairs of correlation functions are in fact
identical.
The obvious topological model
behind this is the $SU_1(3)/SU_1(3)$ model. The fractal
dimension of the contours can be reasonably conjectured to be given
by the
formula
similar to that for the Ising model,
\begin{equation}
h ~=~ {1\over 16}-{1 \over 2} \left({-{1\over 4}}\right) ~=~ {3\over
16} \ ,
\end{equation}
leading to $D_f={13\over 8}$.

More generally one can conjecture similar properties for the contours
in the
high temperature phase of the $Z_{n+1}$ models and
$SU_1(n+1)/SU_1(n+1)$
theories.
The conformal weight $h$ will be $h={c\over 24}-{1\over 2} ({-{c\over
6}})
= {c \over 8}$, where $c = {3n \over n+2}$ is the central charge of
the
$n^{\rm th}$ $N=2$ superconformal minimal model.  This leads to a
predicted
fractal dimension of:
\begin{equation}
D_f ~=~ 2 - {c\over 4} ~=~ {5n+16 \over 4(n+2)} \ .
\end{equation}

\section{Conclusions and open problems}

The subject of $N=2$ lattice models has, as yet, not produced the
same kind of
nice
structures as in the continuum $N=2$  field theories. In particular
it is
important
to emphasize that there is no real algebraic simplification for the
computation of most physical properties.
This suggests that, despite the large number of models exhibited so
far,
some better ones still remain to be discovered. In addition, very
little is
understood
about the ``lattice supersymmetry''
and the way these $N=2$ lattice models provide discretizations of
field theories with fermions. A step in that direction would be to
identify
supercharges on
the lattice and study their anti-commutator in the scaling limit.
This might
be accomplished
in the way that has recently been used for the Virasoro algebra
\cite{WMKHS}.

On the other hand the study of $N=2$ lattice models has already
provided many
interesting
physical ideas, in particular concerning the topological models. We
believe
that
$N=2$ theories are the natural framework to describe most geometrical
statistical
mechanics models and more generally the properties of Bloch walls in
non-critical
spin models. It is also amusing to notice that what is ``physical''
in field
theory and
condensed matter is rather different. For instance spontaneous
breaking of
$N=2$
supersymmetry seems to be a very natural feature in lattice models.
It is
rather
remarkable that the associated level crossings are encoded in the
poles of
differential
equations of Painleve type.

While finding a lattice version of supersymmetry is probably the most
important
issue to resolve, there are many other open problems in the subject
of $N=2$
lattice models.  One can try to establish our observations about
Bloch walls in
generality, or find evidence to support these observations through
numerical
simulations.  As mentionned in section 4, there are unresolved issues
about
supersymmetry and the scatering matrices in the more complex $N=2$
quantum
integrable models.  (Some suggestions about these scattering matrices
have
recently
been made \cite{NewGS}.)  Simple corner transfer matrix computations
have been done for $N=2$ lattice models \cite{DNNW,DNNWnew}.  The
results yield
$N=2$ superconformal characters as one would expect, but the
computation was
done
by the traditional method \cite{Baxter}.  It would be interesting to
see if the
remarkable properties of $N=2$ supersymmetry could be exploited
directly to
reach this result in a much simpler manner.

\index{Knizhnik-Zamolodchikov equations}
There is also the question of
more general lattice correlation functions.   In the continuum limit,
at
criticality, one can obtain Knizhnik-Zamolodchikov equations for the
correlation functions in $N=2$ superconformal coset models.
At first sight, these equations are no simpler than those for
non-super\-symmetric coset models.  However, once one puts in the
special
parameter values for the correlators of chiral primary fields one
finds
that one has saturated some bound and the solution of the
differential equation
is particularly simple.  For example, if the equation is
hypergeometric, the
correlator of chiral primaries is usually one of the special rational
solutions.
It would be most interesting to see if similar simplifications occur
in the
$q$-deformed Knizhnik-Zamolodchikov equations for the lattice
correlators
of \cite{latcor}.

Much of the motivation for these open problems is simply to find some
$N=2$
supersymmetry miracles in exactly solvable lattice models.  We
therefore
leave it as a general challenge to find a another formulation of
these models,
perhaps using the \LG structure more directly, in which the $N=2$
supersymmetry leads to dramatic simplifications.

\vskip3cm
\leftline{\bf Acknowledgements}

One of us (N.~W.) would like thank the organizers of the workshop on
``New
Developments in String Theory, Conformal Models and Topological Field
Theory,''
for their invitation to speak and for creating a very stimulating
workshop.
The authors also wish to thank the organizers for providing them
with
the opportunity to write  this review, which is the result of many
discussions.
The authors wish to thank warmly their  collaborators,  in
particular Paul Fendley, Dennis Nemeschansky, Alyosha Zamolodchikov
and Ziad
Maassarani.
This work was partially supported by the DOE under grant No.
DE-FG03-84ER40168.
H.~S. was also supported by the Packard Foundation.

\vfill
\eject





%



\begin{thebibliography}{100}
\bibitem{Martrev} E.~Martinec, {\it ``Criticality, catastrophes and
compactifications,''}  in the V.G.~Knizhnik memorial volume, L.~Brink
{\it  et
al.}
(editors): {\it ``Physics and mathematics of strings.''}
\bibitem{Wrevs}N.P. Warner, {\it Lectures on $N\!=\!2$

superconformal theories and
singularity theory''},in {\it ``Superstrings '89,''} proceedings of
the Trieste
Spring School, 3--14 April 1989.  Editors: M.\ Green, R.\ Iengo,  S.\
Randjbar-Daemi, E.\ Sezgin and A.\ Strominger.  World Scientific
(1990);
{\it ``$N\!=\! 2$ Supersymmetric Integrable Models and

Topological Field Theories,''}
Lectures given at the Summer School on High Energy Physics and
Cosmology,
Trieste, Italy, June 15th -- July 3rd, 1992.  To appear in the
proceedings.
\bibitem{Vafrev} C.~Vafa, Lectures given at the Summer School on High
Energy
Physics and Cosmology, Trieste, Italy, June 15th -- July 3rd, 1992.
To appear
in the proceedings.
\bibitem{VWMLG} E.~Martinec, \plt{217B} (1989) 431; C.~Vafa and
N.P.~Warner,
\plt{218B} (1989) 51.
\bibitem{LVW}W.~Lerche, C.~Vafa and N.P.~Warner, {\sl Nucl. Phys.}
{\bf B324} (1989) 427.
\bibitem{CIZ}A.~Cappelli, C.~Itzykson and J.B.~Zuber, {\sl Nucl.
Phys.}
{\bf B280} (1987) 445.
\bibitem{EWell} E.~Witten, \nup{403} (1993) 159;
``On the Landau-Ginzburg description of $N=2$ minimal models,''
IASSNS-HEP-93-10, hep-th/9304026.
\bibitem{PDOASY}  P.~di~Francesco, O.~Aharony and S.~Yankielowicz,
``Elliptic Genera and the Landau-Ginzburg approach to $N=2$
orbifolds,''
SACLAY-SPHT-93-068, hep-th/9306157.
\bibitem{TKYYSKY} T.~Kawai, Y.~Yamada and S.-K.~Yang, ``Elliptic
genera and $N=2$ superconformal field theory,'' KEK-TH-362,
hep-th/9306096.
\bibitem{MHenn} M.~Henningson, ``$N=2$ gauged WZW models and the
elliptic
genus,''
IASSNS-HEP-93-39, hep-th/9307040.
\bibitem{FMVW}  P.\ Fendley, S.\ Mathur, C.\ Vafa and N.P. \ Warner,
\plt{243B} (1990) 257.
\bibitem{PFKI} P.~Fendley and K.~Intriligator, \nup{372} (1992) 533;
\nup{380}
(1992) 265.
\bibitem{ALDNNW}A.~LeClair, D.~Nemeschansky and N.P.~Warner \nup{390}
(1993)
653.
\bibitem{CecVaf}S.~Cecotti and C.~Vafa, {\sl Nucl. Phys.} {\bf 367}
(1991) 359.
\bibitem{CFIV}S.~Cecotti, P.~Fendley, K.~Intriligator and C.~Vafa,
{\sl Nucl.
Phys.}
 {\bf B386} (1992) 405.
\bibitem{VPasq} V.~Pasquier, \nup{295} (1988) 491.
\bibitem{VPHS}V.~Pasquier and H.~Saleur, {\sl Nucl. Phys.} {\bf B330}
(1990)
523.
\bibitem{BNI}B.~Nienhuis, {\sl J. Stat. Phys.} {\bf 34} (1984) 731.
\bibitem{DFSZI}P.~di~Francseco, H.~Saleur and J.-B.~Zuber, {\sl J.
Stat. Phys.}
{\bf 49}
(1987)  57.
\bibitem{CoulombG}
P.~Di~Francesco, H.~Saleur and J.-B.~Zuber, \nup{285} (1987) 454;
\nup{300}
(1988) 393;
I.~Kostov, \nup{300} (1988) 559.
J.-B.~ Zuber, ``Conformal Field Theories, Coulomb Gas Picture
and Integrable models'', in {\it Fields, Strings and Critical
Phenomena},
editors E.~Br\'ezin and J.~Zinn-Justin, Les Houches 1988 Session
XLIX.
\bibitem{Kyoto}E.~Date, M.~Jimbo, T.~Miwa and M.~Okada, {\sl Phys.
Rev.}
{\bf B35} (1987) 2105.
\bibitem{FQS}D.~Friedan, Z.~Qiu  and S.~Shenker, {\sl Phys. Lett.}
{\bf 151B}
(1984) 1575.
\bibitem{KMS} D.~Kastor, E.~Martinec and S.~Shenker, {\sl Nucl.
Phys.}
{\bf B316} (1989) 590.
\bibitem{ZQiu}Z.~Qiu, {\sl Nucl. Phys.} {\bf B270} (1986) 205.
\bibitem{OFoda}O.~Foda, {\sl Nucl. Phys.} {\bf B300} (1988) 611.
\bibitem{SKY}S.-K.~Yang, {\sl Nucl. Phys.} {\bf B285} (1987) 183.
\bibitem{SKYHBZ}S.-K.~Yang and H.B.Zheng, {\sl Nucl. Phys.} {\bf
B285} (1987)
410.
\bibitem{Bonn}M.~Baake and G.~von~Gehlen, V.Rittenberg, {\sl
J.~Phys.} {\bf
A20}
(1987) L479; {\sl J.~Phys.} {\bf A20} (1987) L487.
\bibitem{DFSZ}P.~di~Francesco, H.~Saleur and J.-B.~Zuber, {\sl Nucl.
Phys.}
{\bf B300} (1988) 393.
\bibitem{SKYI}S.-K.~Yang, {\sl Phys. Lett.} {\bf B209} (1988) 242.
\bibitem{HSAA}H.~Saleur, {\sl Nucl. Phys.} {\bf B382}(1992) 486, 532.
\bibitem{AGIVEK}A.G.~Izergin and V.E.~Korepin, {\sl Commun. Math.
Phys.} {\bf
79} (1981) 303.
\bibitem{WBNN}S.O.~Warnaar, M.T.~Batchelor and B.~Nienhuis, {\sl J.
Phys.}
{\bf A25} (1992) 3077.
\bibitem{PFHS}P.~Fendley and H.~Saleur, {\sl Nucl. Phys.} {\bf B388}
(1992)
609.
\bibitem{CGGS}C.~Gomez and G.~Sierra, ``Spin anisotropy commensurable
chains:
quantum
 group symmetries and $N=2$ Susy,'' {\sl Nucl.~Phys.} {\bf B}, to
appear.
\bibitem{MNW}Z.~Maassarani, D.~Nemeschansky and N.P.~Warner, {\sl
Nucl. Phys. }
{\bf B393} (1993) 523.
\bibitem{ZMa} Z.~Maassarani, ``Conformal Weights via Bethe Ansatz for
$N=2$
Superconformal Theories,'' USC preprint USC-93/019.
\bibitem{DNNW}D.~Nemeschansky and N.P.~Warner, ``Off-Critical Lattice
Analogues
of $N=2$
Supersymmetric Quantum Integrable Models,'' preprint USC-93/018,
to appear in {\sl Nucl. Phys. B}.
\bibitem{KastFort}P.~Kasteleyn and C.~Fortuin, {\sl J.Phys. Soc.
Japan Suppl.}
{\bf 26} (1969) 11.
\bibitem{BlotNight}H.W.J.~Blote and M.P.M.~Nightingale, {\sl Physica}
{\bf
112A}
(1982)  405.
\bibitem{Baxter}R.J.~Baxter, {\sl ``Exactly solved models in
statistical
mechanics,''}
Academic Press, London 1982.
\bibitem{PMar}P.~Martin, {\sl ``Potts models,''} World Scientific.
\bibitem{MartSal}P.~Martin and H.~Saleur,``The blob algebra and the
periodic
Temperley-Lieb
algebra'', preprint USC-93-009, to appear in {\sl Lett. Math. Phys.}.
\bibitem{Felder} G.~Felder, \nup{317} (1989), 215.
\bibitem{PLKACB}L.P.~Kadanoff and A.C.~Brown, {\sl Ann. Phys. } {\bf
121}
(1979) 318.
\bibitem{VPasqb} V.~Pasquier, \cmp{118} (1988) 335.
\bibitem{HSJBZ} H.~Saleur and J.-B. Zuber, ``Integrable Lattice
Models and
Quantum Groups,''
in the proceedings of the 1990 Trieste Spring School on String Theory
and
Quantum Gravity.
\bibitem{ABF} G.E.~Andrews,  R.J.~Baxter and P.J.~Forrester,
\jsp{35} (1984) 193.
\bibitem{CycSOS}A.~Kuniba and T.~Yajima, \jsp{52} (1987) 829;
P.~Ginsparg, \nup{295} (1988) 153;
P.A.~Pierce and K.~Seaton, \annp{193} (1989) 326;
P.A.~Pierce and M.T.~Batchelor, \jsp{60} (1990) 77.
\bibitem{PMMAW} P.~Mathieu and M.A.~Walton, \plt{254B} (1991) 106.
\bibitem{WLNW} W.~Lerche and N.P.~Warner, \nup{358} (1991) 571.
\bibitem{Grisaru} M.T.~Grisaru, S.~Penati and D.~Zanon, \plt{253B}
(1991) 357;
G.W.~Delius, M.T.~Grisaru, S.~Penati and D.~Zanon, \plt{256B} (1991)
164;
\nup{359} (1991) 125.
\bibitem{Hollow}  J.~Evans and T.J.~Hollowood \nup{352} (1991) 723;
\nup{382} (1992) 662; \plt{293B} (1992) 100.
\bibitem{GFAL} G.~Felder and A.~LeClair, \ijmp{7} (1992) 239.
\bibitem{MATHFRIENDS}F.~M.~Goodman, P.~de la Harpe, V.~F.~R.~Jones,
{\it
``Coxeter
graphs and towers of algebras''}, MSRI Publications  number 14,
Springer
Verlag, and
 references therein.
\bibitem{WMKHS}W.M.~Koo and H.~Saleur, ``Representations of the
Virasoro
algebra from lattice
 models,'' preprint USC-93-025.
\bibitem{MTMS}M.~Takahashi and M.~Suzuki, {\sl Prog.Th.Phys.} {\bf
48} (1972)
2187.
\bibitem{Zamo}Al.B.~Zamolodchikov, {\sl Nucl. Phys.} {\bf B366}
(1991) 122;
{\sl Nucl. Phys.}
 {\bf B358} (1991) 524.
\bibitem{NRHS}N.~Yu~Reshetikhin and H.~Saleur,``Lattice
regularization of
massive and massless
integrable field theories'',  preprint USC-93/-20.
\bibitem{LDFLAT} L.D.~Faddeev and L.A.~Takhtajan, {\sl Phys.Lett.}
{\bf A85}
(1981) 375.
\bibitem{FSrev} P.~Fendley and H.~Saleur, ``Massless integrable QFT
and
massless
scattering in $1+1$ dimensions,'' preprint USC-93-022.
\bibitem{DG}P.G.~de~Gennes, {\it ``Scaling concepts in polymer
physics''},
Cornell
University Press.
\bibitem{FSZ}P.~Fendley, H.~Saleur and Al.B.~Zamolodchikov,
``Massless flows I:
the sine-Gordon and $O(n)$ models'' and ``Massless flows II: the
exact S-matrix
approach'',
{\sl Int. J. Mod. Phys.} {\bf A}, to appear.
\bibitem{kyotoa} M.~Jimbo, T.~Miwa and M.~Okado, \cmp{116} (1988)
507.
\bibitem{BDHS}B.~Duplantier and H.~Saleur, {\sl Nucl. Phys.} {\bf
B290} (1987)
291.
\bibitem{KazSuz} Y.~Kazama and H.~Suzuki, \nup{234} (1989) 73.
\bibitem{NWtopG}D.~Nemeschansky and N.P.~Warner, \nup{380} (1992)
241.
\bibitem{FLMW} P.~Fendley, W.~Lerche, S.D.~Mathur and N.P.~Warner,
\nup{348}
(1991) 66.
\bibitem{WitTop} E.~Witten, \cmp{117} (1988) 353; \cmp{118} (1988)
411;
\nup{340} (1990) 281.
\bibitem{TESKY} T.~Eguchi and S.-K.~Yang, \mpl4 (1990) 1693.
\bibitem{kyotob} T.~Miwa, M.~ Jimbo and M.~Okado, ``Symmetric tensors
of the
$A^{(1)}_{n-1}$ family,'' Kyoto preprint (1987), in {\it Algebraic
Analysis,}
Festschrift for M.~Sato's 60th birthday, Academic Press (1988).
\bibitem{MPRCAT}M.P.~Richey and C.A.~Tracy, \jsp{42} (1986) 311.
\bibitem{kyotoc} E.~Date, M.~Jimbo, A.~Kuniba, T.~Miwa and M.~Okado,
{\it Advanced Studies in Pure Mathematics,} {\bf 16} (1988) 17.
\bibitem{TRKEM} T.R.~Klassen and E.~Melzer, \nup{338} (1990) 485.
\bibitem{CADBAL} D.~Bernard and A.~LeClair,  \plt{247} (1990) 309;
C.~Ahn, D.~Bernard, and A.~LeClair, \nup{346} (1990) 409;
D. Bernard and A. LeClair, \cmp{142} (1991) 99.
\bibitem{THQG} T. Hollowood, ``A Quantum Group Approach to
Constructing
Factorizable S-Matrices,'' Oxford Preprint OUTP-90-15P, June 1990.
\bibitem{TNHJVVF} T.~Nakatsu,  \nup{356} (1991) 499;
H.J.~de~Vega and V.A.~Fateev, \ijmp{6} (1991) 3221.
\bibitem{PCGM} P.~Christe and G.~Mussardo, \nup{330} (1990) 465;
\ijmp{5} (1990) 4581;  G.~Mussardo, \ijmpb{6} (1992) 2061.
\bibitem{EOPW} M. Karowski and H. J. Thun, \nup{190} (1981) 61;
E. Ogievetsky and P. Wiegmann, \plt{168} (1986),  360;
E.~Ogievetsky, N.Yu.~Reshetikhin and P.~Wiegmann, \nup{280} (1987)
45.
\bibitem{BazRes}V.V.~Bazhanov, N.Yu.~Reshetikhin, {\sl Progress in
Theor.
Phys.}
{\bf 102} (1990) 301.
\bibitem{NewGS} C.~Gomez and G.~Sierra, ``On the integrability of
$N=2$
Landau-Ginzburg
models: a graph generalization of the Yang-Baxter equation,'' CERN
preprint
TH-6963/93, hep-th/9309007.
\bibitem{DNNWnew}D.~Nemeschansky and N.P.~Warner, in preparation.
\bibitem{latcor} M.~Jimbo,  K.~Miki, T.~Miwa and A.~Nakayashiki,
``Correlation functions of the $XXZ$-model for $\Delta < -1$,''
kyoto PRINT-92-0191, hep-th/9205055; M.~Jimbo,  T.~Miwa and
A.~Nakayashiki,
\jphys{26} (1993) 2199; O.~Foda,  M.~Jimbo, T.~Miwa,
K.~Miki and A.~Nakayashiki, ``Vertex operators in solvable lattice
models,''
RIMS-922, hep-th/9305100;  M.~Jimbo, T.~Kojima, T.~Miwa, Y.-H.~Quano,
``Smirnov's integrals and quantum Knizhnik-Zamolodchikov equation
of level $0$,'' RIMS-945, hep-th/9309118.
\end{thebibliography}
\end{document}